\documentclass[11pt]{article}
\usepackage[utf8]{inputenc}
\usepackage{url} 
\usepackage{lineno}
\usepackage{graphicx}
\usepackage{hyperref}
\usepackage[affil-it]{authblk}
\usepackage{soul}
\usepackage{amsfonts}  
\usepackage{booktabs}  
\usepackage[margin=1in]{geometry}

\usepackage{amsmath}

\usepackage{natbib}

\begin{document}
\title{\hrule \vspace{15pt}\textbf{WeatherBench 2: A benchmark for the next generation of data-driven global weather models} \vspace{15pt} \hrule}

\author[1,*]{Stephan Rasp}
\author[1]{Stephan Hoyer}
\author[1]{Alexander Merose}
\author[1]{Ian Langmore}
\author[2]{Peter Battaglia}
\author[1]{Tyler Russell}
\author[2]{Alvaro Sanchez-Gonzalez}
\author[1]{Vivian Yang}
\author[1]{Rob Carver}
\author[1]{Shreya Agrawal}
\author[3]{Matthew Chantry}
\author[3]{Zied Ben Bouallegue}
\author[3]{Peter Dueben}
\author[1]{Carla Bromberg}
\author[1]{Jared Sisk}
\author[1]{Luke Barrington}
\author[1]{Aaron Bell}
\author[1]{Fei Sha}

\affil[1]{Google Research}
\affil[2]{Google DeepMind}
\affil[3]{European Centre for Medium-Range Weather Forecasts}
\affil[*]{Corresponding author: srasp@google.com}
\date{}

\maketitle

\begin{abstract}
WeatherBench 2 is an update to the global, medium-range (1--14 day) weather forecasting benchmark proposed by \cite{rasp_weatherbench_2020}, designed with the aim to accelerate progress in data-driven weather modeling. WeatherBench 2 consists of an open-source evaluation framework, publicly available training, ground truth and baseline data as well as a continuously updated website with the latest metrics and state-of-the-art models: \url{https://sites.research.google/weatherbench}. This paper describes the design principles of the evaluation framework and presents results for current state-of-the-art physical and data-driven weather models. The metrics are based on established practices for evaluating weather forecasts at leading operational weather centers. We define a set of headline scores to provide an overview of model performance. In addition, we also discuss caveats in the current evaluation setup and challenges for the future of data-driven weather forecasting.
\end{abstract}

\begin{center}
    \line(1,0){250}
\end{center}

\section{Introduction}

Global, medium-range (1--14 day) numerical weather prediction (NWP) is a key component of modern weather forecasting \citep{bauer_quiet_2015}. It has huge economic and societal impact as many significant weather events occur on this time scale, such as heat waves, tropical- and extra-tropical cyclones, droughts or heavy precipitation leading to flooding. In addition to providing valuable forecasts, global NWP models serve a range of additional purposes: they provide boundary conditions for regional, high-resolution models; they are used to create (re-)analyses; and they serve researchers as tools to better understand the atmosphere. Current NWP models are based on discretizations of the governing equations describing fluid flow and thermodynamics \citep{kalnay_atmospheric_2002}. As of 2023, most global models have a horizontal grid spacing of less than 25\,km. The European Center for Medium-range Weather Forecasting's (ECMWF) Integrated Forecast System (IFS) model, for example, has a resolution of 0.1$^{\circ}$, around 9km, in its high-resolution (HRES) and, since the 2023 upgrade, also in its ensemble (ENS) configuration. This still leaves many important physical processes unresolved, for example cloud physics and radiation. The effect of these processes on the resolved scales has to be approximated in so-called parameterizations \citep{stensrud_parameterization_2007}. Another important aspect of NWP is estimating the current state of the atmosphere, required to initialize model forecasts. This is done using data assimilation algorithms that combine model forecasts and observations to produce an analysis, a best guess of the current state of the atmosphere. The last few decades have seen a steady improvement in global NWP driven by increased computing power, which in turn allowed higher resolution and more ensemble members, better observations and data assimilation, and better representations of the physical processes \citep{magnusson_factors_2013}. Despite the impressive progress in physical global NWP, however, there is still ample room for improvement. Recent studies estimate that the intrinsic limit of predictability of mid-latitude weather is at around 15 days, while the current practical limit of predictability is at around 10 days. Around half of the remaining 5 days of potential skill stem from model improvements, the other half from better initial conditions \citep{zhang_what_2019, selz_transition_2022}.

In the last few years, spurred by the rise of artificial intelligence (AI) in domains like computer vision and natural language processing \citep{lecun_deep_2015}, researchers have been exploring the possibility of using modern machine learning algorithms for weather forecasting. A number of initial attempts at building data-driven medium-range NWP models \citep{dueben_challenges_2018, scher_toward_2018,weyn_can_2019} led to the creation of WeatherBench \cite[from here on called WB1]{rasp_weatherbench_2020}, a benchmark for global, medium-range weather prediction. The goal of WB1 was to provide a common, reproducible framework for evaluating global, data-driven forecasts and compare them against traditional baselines. Benchmarks have been hugely influential in the ML community to measure progress on specific tasks. Famous examples include ImageNet \citep{deng_imagenet_2009} which helped kick-start the AI revolution in computer vision \citep{krizhevsky_imagenet_2017} and the GLUE benchmark \citep{wang_glue_2018} in language. After its release, WB1 was used by a number of studies to explore different machine learning approaches. \cite{weyn_improving_2020} used a cubed sphere projection in combination with a UNet architecture to iteratively predict the atmospheric state at 2$^{\circ}$ resolution. \cite{rasp_data-driven_2021} used a deep Resnet architecture to directly predict fields up to 5 days ahead at a 5.625$^{\circ}$ resolution. \cite{clare_combining_2021} similarly used a Resnet but added a probabilistic output layer. A probabilistic extension to WeatherBench along with several ML baselines was published as well \citep{garg_weatherbench_2022}. An up-to-date leaderboard of WB1 metrics can be found at \url{https://github.com/pangeo-data/WeatherBench}. These studies, all using comparatively coarse resolution, did not come close to the skill of current physical NWP models.


In early 2022, improvements in data-driven weather models considerably picked up in pace. \cite{keisler_forecasting_2022} used a graph neural network (GNN; \cite{pfaff_learning_2021}) iteratively with a time step of 6 hours at a resolution of 1$^{\circ}$ and 13 vertical levels. On deterministic upper-level verification metrics the model achieves skill comparable to some operational NWP models. \cite{pathak_fourcastnet_2022} used a modified vision transformer \citep{guibas_adaptive_2022}, called FourCastNet, for prediction at a very high resolution of 0.25$^{\circ}$ and a 6 hour time step. An updated version using spherical fourier neural operators was also published recently \citep{bonev_spherical_2023}. Pangu-Weather \citep{bi_accurate_2023}, based on a different variation of vision transformers at equally high-resolution, obtained deterministic metrics that outperform HRES. Shortly after, GraphCast \citep{lam_learning_2023} built upon the work of \cite{keisler_forecasting_2022} and scaled a GNN to 0.25$^{\circ}$ horizontal resolution. On many deterministic metrics as well as extreme weather indicators, GraphCast also outperforms HRES. In 2023, another vision transformer variant called FengWu \citep{chen_fengwu_2023} was published with state-of-the-art, deterministic scores for longer forecast lead times. Similarly, the FuXi model \citep{chen_fuxi_2023} achieves deterministic scores similar to the IFS ensemble mean up to 15 days. SwinRDM \citep{chen_swinrdm_2023} combines a recurrent network with a diffusion model to provide granular predictions at 0.25$^{\circ}$ resolution. ClimaX \citep{nguyen_climax_2023} and Stormer \citep{nguyen_scaling_2023} are examples of models achieving competitive skill at lower resolution. NeuralGCM \citep{kochkov_neural_2023} represents the first hybrid ML-physics model that achieves state-of-the-art scores, including an ensemble version.

In light of the rapid advances of the field, the need for an updated benchmark that allows for easy comparison between different approaches becomes apparent. In comparison to WB1, WB2 supports much higher-resolution data and evaluation and adds additional metrics. This paper will lay out the design principles for WB2, followed by a detailed description of the evaluation metrics and datasets, as well as results on the headline scores. We will also discuss several issues with WB2 and potential future directions for this benchmark dataset.

\section{Design decisions for WeatherBench 2}
\label{sec:design}
A general challenge of designing benchmarks for weather prediction is that weather is a very high-dimensional and multi-faceted problem \citep{dueben_challenges_2022}. Every use case has slightly different requirements and quality measures. Take as an example model development at ECMWF, one of the world's leading operational NWP centers. ECMWF tracks a large number of metrics to evaluate their model performance. The headline scores (\url{https://www.ecmwf.int/en/forecasts/quality-our-forecasts}) and scorecard (\url{https://sites.ecmwf.int/ifs/scorecards/}) serve as concise summaries but are only the tip of the iceberg. In addition to quantitative scores, feedback is also be collected from in-house experts and end users, each of them interested in specific aspects of the forecast system. Recognizing the limitations of particular scores is even more important to keep in mind when evaluating ML-based approaches, which often violate standard assumptions in traditional NWP. All this is to say that no single metric or set of metrics will ever be able to fully describe what a "good" forecast is. WB2 does not attempt to do so. Similarly to ECMWF, WB2 defines a set of headline scores (described in Section~\ref{sec:results}) and evaluation tools that aim to capture key aspects of medium-range weather forecasting but are not meant to be exhaustive. Therefore, WB2 should not be seen as a traditional benchmark challenge with a single leaderboard but rather as a tool to compare different approaches on different aspects. 

In its details, the WB2 evaluation protocol sticks closely to the forecast verification used by the WMO \citep{world_meteorological_organization_wmo_manual_2019} and operational weather centers like ECMWF. WB2 does not try to reinvent these standard verification protocols but rather aims to make them accessible to the community interested in AI models. This is done by providing open-source ground truth data and evaluation code, as well as by providing a single place to compare different traditional and AI approaches.

While WB2 defines the targets to be evaluated, it leaves the remaining modeling setup open (e.g., which inputs or model resolution to use). In this sense, WB2 is a benchmark for entire forecast systems, i.e., the combination of choice of input data, training setup and model architecture, rather than focusing on the models specifically. While this makes it harder to compare different model architectures, we believe that designing the non-model part of the forecast system components is equally as important and different approaches should be encouraged to accelerate progress in the field, as long as overfitting is avoided. However, it is important to recognize that different design choices can lead to advantages and disadvantages. One important choice is which input dataset to use for starting ("initializing") forecasts. Currently, most models use re-analysis datasets, in particular ERA5, which would not be available in an operational setup and provide an potential advantage over operational initial conditions (more details in Section~\ref{sec:discussion_ic}).

Another priority of WB2 is to emphasize the importance of probabilistic prediction. Weather forecasting is an inherently uncertain endeavor because of chaotic error growth \citep{lorenz_deterministic_1963, zhang_mesoscale_2007}, implying that even with perfect models and near perfect initial conditions there is a range of possible outcomes. In operational NWP, this fact led to the development of ensemble prediction systems in the 1990s \citep{palmer_ensemble_1993, toth_ensemble_1993}. In an ensemble, a number of forecasts (typically 10--100) are run with perturbed initial conditions and sometimes model physics, providing different potential realizations of future weather. One of the most important advantages of ensemble forecasts is that they can provide reliable information for decision making. For example, they can be used to estimate the probability of extreme events that a single, deterministic forecast might miss. Tropical cyclone track forecasts (often displayed as plumes) are a prime example of this. The data-driven models discussed above, with the exception of a perturbed ensemble experiment in the Pangu-Weather paper, all only provide deterministic forecasts. Similarly to the evolution of dynamical forecast systems, data-driven forecasts will need to become probabilistic to provide users with the most actionable information. This could be possible as a post-processing step on top of deterministic models, for example, through ensemble dressing or lagged forecasts. However, the success of NWP ensembles hints at the advantages of designing probabilistic forecast systems from the ground up, especially for capturing spatio-temporal correlations. For this reason, WB2 will include operational probabilistic verification metrics and baselines from the start. In data-driven modeling, there are several ways to create probabilistic forecasts, from predicting parameterized marginal distributions directly \citep{andrychowicz_deep_2023} to producing generative roll-outs. Some aspects of how to evaluate these forecasts are discussed in Section~\ref{sec:discussion_realism}.

Finally, WB2 will provide a dynamic, open-source framework that can evolve with the needs of the ML-weather community. All data and code needed for evaluation are publicly available and can be extended to accommodate more detailed evaluation in the future, driven either by us or community contributors.

\section{Data, baselines and data-driven models}
\label{sec:data}
Most of the datasets below are available in on Google Cloud Storage in Zarr format. For the latest details on these datasets, please visit \url{https://weatherbench2.readthedocs.io/en/latest/data-guide.html}. Table~\ref{tab:models} provides a summary of key facts for each of the datasets/models, including the training resources.

\begin{table*}[]\centering
\scriptsize
\centerline{
\begin{tabular}{p{2.5cm}p{1.6cm}p{1.8cm}p{0.9cm}p{1.1cm}p{1.8cm}p{1.8cm}p{1.8cm}}
\toprule
\textit{\textbf{Model/ Dataset}} & \textbf{Type} & \textbf{Initial conditions} & \textbf{$\Delta x$} & \textbf{Levels} & \textbf{Training data}    & \textbf{Training resources} & \textbf{Inference time}     \\
\midrule
\textit{ERA5}                   & Reanalysis    &                             & 0.25$^{\circ}$                                     & 137                                                                           &                           &                             &                          \\
\textit{IFS HRES}               & Forecast      & Operational                 & 0.1$^{\circ}$                                      & 137                                                                          &                           &                             &   $\sim$ 50 minutes (*)                       \\
\textit{IFS ENS}                & Forecast      & Operational                 & 0.2$^{\circ}$                                      & 137                                                                      &                           &                             &                          \\
\textit{ERA5 forecasts}         & Hindcast      & ERA5                        & 0.25$^{\circ}$                                     & 137                                                                       &                           &                             &                         \\
\textit{Keisler (2022)}         & Forecast      & ERA5                        & 1$^{\circ}$                                        & 13                                                                         & ERA5 (35 years, see text) & 5.5 days; one A100 GPU      & $\sim$1 second; single GPU  \\
\textit{Pangu-Weather}          & Forecast      & ERA5                        & 0.25$^{\circ}$                                     & 13                                                                    & ERA5 (1979-2017)          & 16 days; 192 V100 GPUs      & several seconds; single GPU \\
\textit{GraphCast}              & Forecast      & ERA5                        & 0.25$^{\circ}$                                     & 37                                                                        & ERA5 (1979-2019)          & 4 weeks; 32 TPU v4          & $\sim$1 minute; single TPU \\
\textit{FuXi}              & Forecast      & ERA5                        & 0.25$^{\circ}$                                     & 13                                                                        & ERA5 (1979-2017)          & $\sim$8 days; 8 A100 GPUs          &  \\
\textit{SphericalCNN}              & Forecast      & ERA5                        & 1.4x0.7$^{\circ}$                                     & 7                                                                        & ERA5 (1979-2017)          & 4.5 days; 16 TPU v4          & \\
\textit{NeuralGCM 0.7$^{\circ}$}              & Forecast      & ERA5                        & 0.7$^{\circ}$                                     & 32                                                                        & ERA5 (1979-2017)          & 3 weeks; 256 TPU v5          & $\sim$1 minute; single TPU \\
\textit{NeuralGCM ENS}              & Forecast      & ERA5                        & 1.4$^{\circ}$                                     & 32                                                                        & ERA5 (1979-2017)          & 10 days; 128 TPU v5          & $\sim$1 minute; single TPU \\
\bottomrule
\end{tabular}}
\caption{Table of datasets and models; $\Delta x$ refers to the horizontal resolution and ``levels'' to the number of vertical model levels used in the input and output. (*) See details in the text for IFS inference time. GPU = Graphics Processing Unit. TPU = Tensor Processing Unit.}
\label{tab:models}
\end{table*}

\subsection{ERA5}
\label{sec:data_era5}
The ERA5 dataset \citep{hersbach_era5_2020} is used as the ground truth dataset for WB2 and as the training dataset for many of the data-driven approaches described above. ERA5 is a reanalysis dataset based on a 0.25$^{\circ}$ (roughly 30\,km) version of ECWMF's HRES model operational in 2016 (cycle 42r1) and ECMWF's 4D-Var data assimilation which uses a wide range of direct and remote sensing observations. ERA5 uses 12 hour assimilation windows from 21--09 and 09-21 UTC, during which a ``prior'' forecast initialized from the previous assimilation window is combined with observations to produce an analysis, a ``best guess'' of the Earth system state during this window. ERA5 data is available at hourly resolution from 1940 to present at the Copernicus Climate Data Store (\url{https://cds.climate.copernicus.eu/}). A subset of the ERA5 dataset is available on the cloud in the cloud-optimized Zarr format.

Note that using ERA5 for evaluation and training has some caveats. First, while attempting to be close to observations, ERA5 is a model simulation which is closer to the truth for some variables than for others. Especially for precipitation, ERA5 sometimes shows large differences to rain gauge measurements \cite{lavers_evaluation_2022}. We still include precipitation evaluation based on ERA5 here but advise caution when interpreting those results (for more details, see discussion in Section~\ref{sec:discussion_era5}). 
Second, ERA5 uses a longer assimilation window compared to operational forecasts (see below). This helps to better constrain the guess of the atmospheric state but would delay the initialization of forecasts in a real time setting. At 00 UTC, for example, ERA5's assimilation window goes 9 hours "into the future". Operationally, the data assimilation window only extend 3 hours ahead from the forecast initialization time. One could use 06/18UTC initialization to level the playing field a little bit more. This issue has been extensively discussed in the supplement of \cite[Fig. 10]{lam_learning_2023} who found that forecasts initialized from ERA5 at 00/12 UTC indeed perform better than those initialized at 06/18UTC, when the ERA5 analysis also only has a 3 hour look-ahead. The difference in RMSE for GraphCast was up to 5\%. However, the 06/18UTC operational analysis are also produced using fewer DA cycles compared to the 00/12UTC analysis.
For WB2, we chose to evaluate forecasts initialized at 00/12UTC. The main reasons for this are that some key baselines (IFS ENS and the ERA5 forecasts) were only available for 00/12 UTC initializations and that this has been the standard for most other ML-based evaluations so far, which allows us to directly include some of them here.

\subsubsection{ERA5 forecasts}

For research purposes, ECMWF ran a set of 10 day hindcasts initialized from ERA5 states at 00/12UTC with the same IFS model version used to create ERA5. These forecast are available in the MARS archive (MARS parameters: class=ea, stream=oper, expver=11, type=fc). Here we downloaded the variables required to compute the headline scores for the year 2020 (with the exception of total precipitation which was not available). Note that data until 5 days lead time is available at 6h intervals, and 12h intervals from 5 to 10 days lead time.

The ERA5 forecasts provide a like-for-like baseline for an AI model initialized from and evaluated against ERA5. They benefit from the same, longer assimilation window compared to the operational initial conditions and are run at 0.25$^{\circ}$ resolution---similar to many modern AI methods. Because of the lower resolution and older model relative to the operational IFS HRES in 2020, one would expect the operational model to be more skillful by itself.

\subsection{Climatology}
\label{sec:data_clim}
The climatology is used for computing certain skill scores, in particular the anomaly correlation coefficient (ACC) and the stable equitable error in probability space (SEEPS), and as a baseline forecast. Here we follow \cite{jung_scale-dependent_2008} and compute the climatology $c$ as a function of the day of year (\textit{doy}) and the time of day (\textit{tod}) by taking the mean of ERA5 data from 1990 to 2019 (inclusive) for each grid point. A sliding window of 61 days is used around each \textit{doy}-\textit{tod} combination with weights linearly decaying to zero from the center. This removes sample noise and makes the climatology smoother in time, at the expense of reducing the seasonal amplitude.

A probabilistic version of the climatology is created by taking each of the 30 years from 1990 to 2019 as an ensemble member, this time without smoothing.

Note that a 30 year climatology will include some climate drift, especially for temperature. Here, we do not apply any measure to correct for this.

\subsection{IFS HRES}

Our main baseline comes from operational forecasts created with ECMWF's IFS model, which, measured by standard verification metrics, are typically regarded as the best global, medium-range weather forecasts. Since 2016, the IFS in its HRES configuration has been run at 0.1$^{\circ}$ (roughly 9\,km) horizontal resolution. The operational model is updated regularly, approximately once to twice a year, which means that the exact model configuration might change during the evaluation period. Usually, updates are associated with slight improvements in most evaluation metrics, though not all. However, changes in the IFS are typically gradual. 
A comprehensive model description can be found at \url{https://www.ecmwf.int/en/publications/ifs-documentation}. A schedule of model upgrades can be found at \url{https://confluence.ecmwf.int/display/FCST/Changes+to+the+forecasting+system}. Initial conditions are created every 6 hours using an ensemble 4D-Var system using information from the previous assimilation cycle's forecast as well as observations in a +/- 3 hour window. After accounting for the time to perform data assimilation and forward simulation, forecasts have a latency of 5.75 to 7 hours from the time at which they are initialized (\url{https://confluence.ecmwf.int/display/DAC/Dissemination+schedule}). Forecasts started at 00 and 12 UTC are run up to a lead time of 10 days. 06 and 18 UTC initializations are run for 3.75 days.

Running a 15 day TCO1279 ($\sim$ 9km) simulation in the setup used for the operational ensemble (using the new high-resolution setup) takes around 52 minutes with I/O and 46 minutes without I/O on 64 128-core (AMD EPYC Rome) nodes.


\subsubsection{IFS HRES Initial Conditions}

For evaluating IFS forecasts, we use operational analyses as ground truth rather than ERA5. This is because evaluating IFS forecasts against ERA5 would result in a non-zero error at time step $t=0$ and therefore would put the IFS at an unfair disadvantage compared to data-driven models trained on and evaluated against ERA5. The difference is most pronounced during the early lead times and becomes smaller at longer lead times. See Fig.~S2 for a comparison. Note that for some variables like 2m temperature differences persist even for longer lead times. This is likely due to a bias between the HRES analysis and ERA5.

Here, as in \cite{lam_learning_2023} we use the initial conditions, i.e. the IFS HRES forecasts at t=0 as our ``analysis''. Note that this dataset is slightly different from the official analysis product on the ECMWF archive. This is because for the official analysis product  an additional surface data assimilation step is taken. For ECMWF's internal evaluation, the official analysis product is used. However, the qualitative differences in the evaluation scores resulting from this discrepancy should be small except for the first time steps after initializations and are mostly limited to surface temperature. For very short lead times, results should not be over-interpreted anyway. Note that for precipitation accumulations we use ERA5 as a precipitation ground truth for all models.

\subsection{IFS ENS}

ECMWF also runs an ensemble version (ENS) of IFS, at 0.2$^{\circ}$ resolution until to the 2023 upgrade (including the 2020 evaluation period used here), now at 0.1$^{\circ}$ resolution. The ensemble consists of a control run and 50 perturbed members. The initial conditions of the perturbed members are created by running an ensemble of data assimilations (EDA) in which observation errors, model errors and boundary condition errors are represented by perturbations (\url{https://confluence.ecmwf.int/display/FUG/Section+5.1.1+Ensemble+of+Data+Assimilations+-+EDA}). The difference of each of the EDA analyses to the mean is then used as a perturbation to the HRES initial conditions. In addition, to more accurately represent forecast uncertainty, singular vector perturbations are added (\url{https://confluence.ecmwf.int/display/FUG/Section+5.1.2+Singular+Vectors+-+SV}) to the initial conditions. Model uncertainties during the forecasts are represented by stochastically perturbed parameterization tendencies \cite[SPPT]{buizza_stochastic_1999}, in which spatially and temporally correlated perturbations are added to the model physics tendencies. Ensemble forecasts are initialized at 00 and 12 UTC and run out to 15 days.


\subsubsection{IFS ENS Mean}

We also include the IFS ENS mean as a baseline, which we computed by simply averaging over the 50 members. The ensemble mean does not represent a realistic forecast but often performs very well on deterministic metrics.

\subsection{Keisler (2022) Graph Neural Network}

\cite{keisler_forecasting_2022} used a graph neural network architecture \citep{pfaff_learning_2021} with an encoder that maps the original 1$^{\circ}$ latitude-longitude grid to an icosahedron grid, on which several rounds of message-passing computations are performed, before decoding back into latitude-longitude space. The model takes as input the atmospheric states at $t=0$ and $t=-6h$ and predicts the state at $t=6h$. To forecast longer time horizons, the model's outputs are fed back as inputs autoregressively. The state consists of 6 three-dimensional variables at 13 pressure levels. ERA5 data is used for training with 1991, 2004 and 2017 used for validation, 2012, 2016 and 2020 for testing and the remaining years from 1979 to 2020 for training. During training the model is trained to minimize the cumulative error of up to 12 time steps (3 days). 

\subsection{Pangu-Weather}

Pangu-Weather \citep{bi_accurate_2023} is a data-driven weather model based on a transformer architecture. It predicts the state of the atmosphere at $t=t+\Delta t$ based on the current state. The state is described by 5 upper-air variables and 4 surface variables on a 0.25$^{\circ}$ horizontal grid (same as ERA5) with 13 vertical levels for the upper-air variables. The model is trained using ERA5 data from 1979 to 2017 (incl.) with 2019 for validation and 2018, 2020 and 2021 for testing. Here we evaluate forecasts for 2020. Four different model versions are trained for different prediction time steps $\Delta t = \{1h, 3h, 6h, 24h\}$. To create forecasts for an arbitrary lead time model predictions are chained together autoregressively from the four different lead time models, using the fewest number of steps. For example, to create a 31h forecast, a 24h forecast is followed by a 6h and then a 1h forecast. The maximum lead time for the data used here is 7 days. Inference code for Pangu-Weather can be found at \url{https://github.com/198808xc/Pangu-Weather}.

\subsubsection{Pangu-Weather (operational)}

The scorecard also contains a version of Pangu-Weather initialized with the operational IFS HRES initial conditions (see above).

\subsection{GraphCast}

GraphCast \citep{lam_learning_2023} is similar in structure to \cite{keisler_forecasting_2022} but uses a higher resolution input with 6 upper-level variables on a 0.25$^{\circ}$ horizontal grid with 37 vertical levels, and additionally 5 surface variables. Further, the processor operates on a multi-mesh, i.e. a nested structure of different resolution meshes. The model is also trained autoregressively up to a time horizon of 12 time steps (3 days). Here, we evaluate a version of GraphCast that was trained on ERA5 data from 1979 to 2019 (incl.). See Suppl. 5.1 of \cite{lam_learning_2023} for details. Code for GraphCast can be found at \url{https://github.com/deepmind/graphcast}.

\subsubsection{GraphCast (operational)}

The scorecard also contains a version of GraphCast fine-tuned and initialized with the operational IFS HRES initial conditions (see above). For mode detail, refer to \url{https://github.com/deepmind/graphcast}.

\subsection{FuXi}
FuXi \citep{chen_fuxi_2023} is an autoregressive cascaded ML weather forecast system based on a transformer architecture with specific models trained for short (0-5 days), medium (5-10 days) and long-range (10-15 days) prediction. The model is trained on ERA5 data at 0.25$^{\circ}$ horizontal resolution with 13 vertical levels. 

\subsection{SphericalCNN}
Spherical Convolutional Neural Networks \citep{esteves_scaling_2023} generalize Convolutional Neural Networks (CNNs) to functions on the sphere, by using spherical convolutions as the main linear operation. The weather model produces output at a resolution 1.4$^{\circ}$ zonally and 0.7$^{\circ}$ meridionally, with 7 vertical levels. Code can be found at \url{https://github.com/google-research/spherical-cnn}.

\subsection{NeuralGCM}
Neural General Circulation Models (NeuralGCM) \citep{kochkov_neural_2023} combine a differential dynamical core with learned physics. NeuralGCM has been trained in a high-resolution deterministic version at 0.7$^{\circ}$ and a lower-resolution ensemble version at 1.4$^{\circ}$. Both model versions use 32 vertical levels.

\section{Evaluation protocol and metrics}
\label{sec:evaluation}

The WB2 evaluation protocol sticks closely to the forecast verification used by the WMO \citep{world_meteorological_organization_wmo_manual_2019} and operational weather centers like ECMWF. Table~\ref{tab:notation} provides a list of the notation used.

\begin{table*}[t!]
\label{tab:notation}

\small
\centering
\begin{tabular}{ccc}
\toprule
\textbf{Symbol} & \textbf{Range}         & \textbf{Description}           \\ \midrule
$f$    &               & Forecast              \\
$o$    &               & Ground Truth          \\
$c$    &               & Climatology           \\
$t$    & ${1, ..., T}$ & Verification time     \\
$l$    & ${1, ..., L}$ & Lead time             \\
$i$    & ${1, ..., I}$ & Latitude index        \\
$j$    & ${1, ..., J}$ & Longitude index       \\
$m$    & ${1, ..., M}$ & Ensemble member index \\
\bottomrule
\end{tabular}
\caption{Notation used in the evaluation metrics.}
\end{table*}

\subsection{Evaluation time period and initialization times}

In its initial version WB2 uses the year 2020. 2020 was chosen a) because it provides a compromise between recency and leaving several more recent years for independent testing if modeling groups desire to do so; and b) because data was available for many of the AI baseline models. One year is also a robust enough sample size for most of the metrics defined below. It should be noted, however, that for metrics focused on extreme events, e.g. hurricanes or heat waves, larger sample sizes are required. As WB2 is updated, evaluation can be changed to a more recent or longer period.

It is also worth discussing that many of the AI models were trained with data only up to 2018. Analysis for the GraphCast model shows that training with more recent data is advantageous \citep{lam_learning_2023}. Further, the IFS model is continuously improved, suggesting that more recent forecasts are better. However, it is difficult comparing absolute values, since there are natural fluctuation year-by-year. In Fig.~S1 we show scores for 2018 (also available on the website). The relative differences in scores between 2018 and 2020 are small, implying that, at least for the metrics shown here, the results are robust.

Evaluation is done on all 00 and 12 UTC initialization times for 2020, i.e. from January 1 2020 00UTC to December 31 12 UTC. This means that some forecasts will extend into 2021. We compared this to evaluating only forecast lead times that are valid in 2020 but the difference in the scores was negligible. The evaluation time step can be freely chosen. We use 6 hours as our highest resolution.

\subsection{Evaluation resolution and area}

Before the computation of the metrics, all forecasts and ground truths are first-order conservatively regridded to 1.5$^{\circ}$ resolution. A regridding tool is available on the WB2 GitHub page. 1.5$^{\circ}$ is also used as the standard resolution for evaluation by the WMO \citep{world_meteorological_organization_wmo_manual_2019} and at ECMWF. Regridding to a common lower resolution allows all models to be evaluated without penalizing coarser-resolution models. Absolute metric values can differ significantly between resolutions but the relative differences between different models are consistent across different evaluation resolution (see Fig.~S9 for a comparison of IFS HRES scores at different resolutions and consult the WeatherBench 2 website to see all models evaluated at different resolutions). However, this should not give the false impression that higher-resolution models are not more accurate as relevant small scale details can be resolved.

All results shown in this paper are computed over all global grid points. Additionally, scores computed many other regions are shown on the website: \url{https://sites.research.google/weatherbench}.

\subsubsection{Below-ground grid points}

For some pressure levels (e.g. 850hPa), grid points in high-altitude regions will be ``below'' ground; i.e., the surface pressure is actually smaller than the pressure level. ERA5 and IFS output will still provide interpolated values at these locations, even though they do not correspond to real physical variables. Here, we exclude below ground grid points from the computation of all metrics. To do so, we compute the climatological fraction of above-ground grid points (defined by \textit{geopotential} $>$ \textit{geopotential at surface}) for each pressure level using the same \textit{doy}-\textit{tod} dependent method described in Section~\ref{sec:data_clim}. We then use this fraction as an additional weight when computing spatial averages. We chose to implement the below-ground mask based on climatological values rather than the actual forecast or ground truth values because ML models, when explicitly trained against masked scores conditioned on the forecast or ground truth, could lead to biased predictions.

\subsection{Deterministic metrics}

All metrics are computed using an area-weighting over grid points. This is because on an equiangular latitude-longitude grid, grid cells at the poles have a much smaller area compared to grid cells at the equator. Weighing all cells equally would result in an inordinate bias towards the polar regions. The latitude weights $w(i)$ are computed as:
\begin{equation}
    w(i) = \frac{\sin \theta_i^\text{u} - \sin \theta_i^\text{l}}{ \frac{1}{I} \sum^{I}_i (\sin \theta_i^\text{u} - \sin \theta_i^\text{l})},
\end{equation}
where $\theta^u_i$ and $\theta^l_i$ indicate upper and lower latitude bounds, respectively, for the grid cell with latitude index $i$.
All vertical (pressure) levels are treated separately. For readability, no level index is included in the equations below.

\subsubsection{Root mean squared error (RMSE)}

The RMSE is defined for each variable and level as
\begin{equation}
    \mathrm{RMSE}_l = \sqrt{\frac{1}{T I J} \sum^T_t \sum^I_i \sum^J_j w(i) (f_{t, l, i, j} - o_{t, i, j})^2}
\end{equation}

Note that this agrees with the definition of the RMSE used by the WMO and ECMWF. In WB1, the time mean was taken outside the square root. We compared both versions and found differences to be small (less than 2\% absolute difference on average).

The wind vector (WV) RMSE is computed as 
\begin{equation}
    \mathrm{RMSE}^{\mathrm{WV}}_l = \sqrt{\frac{1}{T I J} \sum^T_t \sum^I_i \sum^J_j w(i) \left[ (u^f_{t, l, i, j} - u^o_{t, i, j})^2 + (v^f_{t, l, i, j} - v^o_{t, i, j})^2\right]}
\end{equation}
where $u^f$, $u^o$, $v^f$ and $v^o$ are the $u$ and $v$ components of wind in the forecast and observations, respectively.

\subsubsection{Anomaly correlation coefficient (ACC)}

The ACC is computed as the Pearson correlation coefficient of the anomalies with respect to the climatology $c$: 
\begin{equation}
    f'_{t, l, i, j} = f_{t, l, i, j} - c_{t, l, i, j}; \quad o'_{t, i, j} = o_{t, i, j} - c_{t, i, j}
\end{equation}
where $c_{t, l, i, j}$ and $c_{t, i, j}$ indicate the climatology corresponding to the appropriate \textit{tod}-\textit{doy} combination. The ACC then is defined as 
\begin{equation}
    \mathrm{ACC}_l = \frac{1}{T}\sum^T_t\frac{
    \sum^I_i \sum^J_j w(i) f'_{t, l, i, j} o'_{t, i, j}
    }{
    \sqrt{
    \sum^I_i \sum^J_j w(i) f'_{t, l, i, j}{^2} \sum^I_i \sum^J_j w(i) o'_{t, i, j}{^2}
    }
    }
\end{equation}
The range of ACC values goes from 1, indicating perfect correlation, to -1, indicating perfect anti-correlation. A climatological forecast has an ACC value of zero. ECMWF states that when ``ACC value falls below 0.6 it is considered that the positioning of synoptic scale features ceases to have value for forecasting purposes'' (\url{https://confluence.ecmwf.int/display/FUG/Anomaly+Correlation+Coefficient}).

\subsubsection{Bias}

The mean error, or simply bias, is computed for each location $i, j$ as 
\begin{equation}
    \mathrm{Bias}_{l, i, j} = \frac{1}{T}\sum^T_t (f_{t, l, i, j} - o_{t, i, j})
\end{equation}
In addition, we compute the globally averaged root mean squared bias (RMSB) as
\begin{equation}
    \mathrm{RMSB}_{l} = \sqrt{\frac{1}{I J} \sum^I_i \sum^J_j w(i) \mathrm{Bias}_{l, i, j}^2}
\end{equation}

\subsubsection{Stable Equitable Error in Probability Space -- SEEPS}

Traditional deterministic scores such as RMSE and ACC are not good choices for evaluating precipitation forecasts. This is because precipitation has a very skewed distribution and high spatio-temporal intermittency or unpredictability. Under such conditions, traditional scores heavily favor unrealistically smooth forecasts. This is the case for all variables but is especially dramatic for skewed variables. For this reason, ECMWF and WMO decided to use the SEEPS score \citep{rodwell_new_2010} for their routine deterministic precipitation evaluation. The SEEPS score is based on a three-class categorization into ``dry'', ``light'' and ``heavy'' precipitation. The score is designed to discourage ``hedging'' (i.e. smooth forecasts) and be stable to parameter choices. For details behind the score, please consult \cite{rodwell_new_2010}. Here, we describe how the score is computed and how the computation differs slightly from that in the original paper.

We use a dry threshold of 0.25\,mm/day (0.1\,mm/day for 6 hourly accumulation; shown on the website). The remaining precipitation values are classified into light and heavy, so that climatologically there are twice as many light compared to heavy precipitation days. We compute this by calculating the 2/3rd quantile of non-dry days for each day-of-year according to the same procedure for computing a smooth climatology described in Section~\ref{sec:data_clim}. This differs from most other SEEPS computations which uses monthly climatologies. Because the daily climatology is smooth though, this does not affect the results much. Forecast/observation pairs are then categorized into the three classes and a 3x3 contingency table is created for each forecast lead time. The contingency table is then multiplied by the scoring matrix $S$ based on the yearly average climatological occurrence of dry days $p_1$ for each geographical location:
\[ 
S = \frac{1}{2}
\begin{bmatrix}
  0 & \frac{1}{1-p_1} & \frac{4}{1-p_1} \\
  \frac{1}{p_1} & 0 & \frac{3}{1-p_1} \\
  \frac{1}{p_1} + \frac{3}{2+p_1} & \frac{3}{2+p_1} & 0
 \end{bmatrix}
\]
where columns represent observed probabilities and rows represent forecast probabilities.

Very wet and very dry regions are excluded. Here we use $0.1 < p_1 < 0.85$ as suggested by \cite{rodwell_new_2010}. Finally an area-weighted average is taken over all locations.

In the figures, we show (1 - SEEPS), the more common skill score version that is positively oriented.

\subsection{Probabilistic metrics}

\subsubsection{Continuous ranked probability score (CRPS)}

Given scalar ground truth $Y$, and i.i.d. predictions $X, X'$, CRPS is defined as $\mathbb{E}|X - Y| - \frac{1}{2}\mathbb{E}|X - X'|$ \citep{gneiting_strictly_2007}. The skill term, $\mathbb{E}|X - Y|$ penalizes poor predictions, while $-\frac{1}{2} \mathbb{E}|X - X'|$ encourages spread. CRPS is minimized just when $X$ is drawn from the same distribution as $Y$. To see this, consider two i.i.d. ground truth samples, $Y, Y'$, then subtract $\frac{1}{2}\mathbb{E}|Y - Y'|$ from CRPS to arrive at the divergence relation
\begin{eqnarray*}
  \mathbb{E}\left|X - Y\right| - \frac{1}{2}\mathbb{E}\left|X - X'\right| - \frac{1}{2}\mathbb{E}\left|Y - Y'\right|
  &=& \int \left( \mathrm{P}[X \leq y] - \mathrm{P}[Y\leq y] \right)^2\,dy.
\end{eqnarray*}
Thus, scalar CRPS is equal, up to a constant independent of the prediction, to the squared L2 difference of their cumulative distribution functions. In the case of a deterministic prediction, the CRPS reduces to the MAE.

WeatherBench considers multi-dimensional predictions, $f_{t,l}$, conditional on the initial ground truth $o_{t-l}$. For these, CRPS is defined by averaging over time/latitude/longitude components. We also take advantage of $M\geq2$ predictions $\{f^{(1)},\ldots,f^{(M)}\}$. Setting
\begin{eqnarray*}
  \|g\|_{t,l} &:=& \frac{1}{IJ} \sum_i^I \sum_j^J w(i) |g_{t,l,i,j}|,
\end{eqnarray*}
we define
\begin{equation}
  \label{equation:crps}
  CRPS_l :=
  \frac{1}{T}\sum_t^T \left[
  \frac{1}{M}\sum_{m=1}^M \|f^{(m)} - o\|_{t,l} - \frac{1}{2M (M-1)}\sum_{m=1}^M\sum_{n=1}^M \|f^{(m)} - f^{(n)}\|_{t,l}
  \right].
\end{equation}
This is the time average of unbiased (conditional) CRPS \citep{zamo_estimation_2018}. With large enough $T$, this is an accurate estimate of CRPS conditional on $o_{t-l}$.
It is therefore minimized by {\em any} prediction $f_{t,l}$ with the same component distributions as ground truth at time $t$, conditional on $o_{t-l}$.

The second term \eqref{equation:crps} is efficiently computed in $O(M\log M)$ time using $O(M)$ memory with a sort rather than a double summation. CRPS for deterministic predictions (where $M=1$) is also supported, and in this case CRPS reduces to (weighted) mean absolute error.


\subsubsection{Spread-skill ratio}

The spread-skill ratio $R$ is defined as the ratio between the ensemble spread and the RMSE of the ensemble mean $\overline{f}_{t, l, i, j} = \frac{1}{M} \sum_m^M f_{t, l, i, j, m}$.

\begin{equation}
    \mathrm{Spread}_l = \sqrt{\frac{1}{T I J} \sum^T_t \sum^I_i \sum^J_j w(i) var_m(f_{t, l, i, j, m})}
\end{equation}
with $var_m$ being the variance in the ensemble dimension.
\begin{equation}
    R = \frac{\mathrm{Spread}}{\mathrm{RMSE}(\overline{f})}
\end{equation}
A well-calibrated ensemble forecast should have a spread-skill ratio of 1 \citep{fortin_why_2014}. Smaller values indicate an underdispersive forecast, while larger values indicate an overdispersive forecast. Note that the spread-skill ratio is only a first-order test for calibration. To further diagnose ensemble calibration, rank histograms are a suitable choice \cite[Ch. 7.7.2]{wilks_statistical_2006}. We plan to include those in WB2 soon.

\subsection{Energy Spectra}

Zonal spectral energy along lines of constant latitude is computed as a function of wavenumber (unitless), frequency (m$^{-1}$) and wavelength (m).

With $f_l$ discrete values along a zonal circle of constant latitude, with circumference $C$, the DFT $F_k$ is computed as
\begin{equation*}
    F_k = \frac{1}{L} \sum_{l=0}^{L-1} f_l e^{-i 2\pi k l/L}.
\end{equation*}
The energy spectrum is then set to
\begin{equation*}
    S_0 := C |F_0|^2,\qquad S_k = 2 C |F_k|^2,\quad k = 1,\ldots, \lfloor L / 2 \rfloor.
\end{equation*}
The factor of 2 appears for wavenumbers $k>0$ since they account for both negative and positive frequency content.

This choice of normalization ensures that Parseval's relation for energy holds:
Supposing $f_l$ are sampled values of continuous function $f(\ell)$, for $0 < \ell < C$ (m), then $(C/L)$ is the spacing of longitudinal samples, whence
\begin{equation*}
    \int_0^C |f(\ell)|^2\,d\ell
    \approx \frac{C}{L} \sum_{l=0}^{L-1} |f_l|^2
    = \sum_{k=0}^{\lfloor L / 2\rfloor} S_k.
\end{equation*}

To arrive at a final spectrum we average the zonal spectra for $30^{\circ} < |lat| < 60^{\circ}$.

\subsection{Headline scores}
The headline scores are listed in Table~\ref{tab:headline}. They reflect the most commonly evaluated variables in medium-range forecasting. The upper-level variables are chosen to capture the large-scale evolution of the atmosphere. Z500 and T850 are particularly good tracers of extra-tropical dynamics. Q700 provides a proxy for moisture transport and, indirectly, clouds. The surface variables are closely aligned with weather impact through 2\,m temperature (T2M), 10\,m wind speed (WS10) and 24\,h precipitation accumulation (TP24hr). The mean sea level pressure (MSLP) is another measure of larger scale dynamics and is a good proxy for the strength of tropical and extra-tropical cyclones. 

\begin{table*}[]
\small
\centering
\begin{tabular}{cccc}
\toprule
\textbf{Variable}        & \textbf{Short name} & \textbf{Deterministic metric} & \textbf{Probabilistic metric} \\ \midrule
\multicolumn{4}{c}{\textit{Upper-level variables}}                                                                      \\
500hPa Geopotential      & Z500                & RMSE                          & CRPS                          \\
850hPa Temperature       & T850                & RMSE                          & CRPS                          \\
700hPa Specific humidity & Q700                & RMSE                          & CRPS                          \\
850hPa Wind vector/speed & WV/WS850            & RMSE (WV)                     & CRPS (WS)                         \\
\midrule
\multicolumn{4}{c}{\textit{Surface variables}}                                                                          \\
2m Temperature           & T2M                 & RMSE                          & CRPS                          \\
10m Wind speed           & WS10                & RMSE                          & CRPS                          \\
Mean sea level pressure  & MSLP                & RMSE                          & CRPS                          \\
24h precipitation        & TP24hr               & SEEPS                         & CRPS                         \\
\bottomrule
\end{tabular}
\caption{List of headline scores. WV is short for Wind Vector; WS is short for Wind Speed.}
\label{tab:headline}

\end{table*}

\section{Results}
\label{sec:results}

The description of the results in this paper will focus on the general characteristics of the evaluation metrics and the baselines, with a smaller focus on comparing the data-driven models themselves. This is because the paper is a snapshot of the state-of-the-art of data-driven modeling at the time of writing and will almost certainly be outdated soon. For this reason, and because visualization of all models on a single graph makes it unreadable, we only show detailed scores for a smaller selection of models. Those models (GraphCast, Pangu-Weather and NeuralGCM) were chosen because of their current relevance in the community. For an up-to-date view of the data-driven state-of-the-art, including detailed scores of all models, visit the accompanying website: \url{https://sites.research.google/weatherbench}.

\subsection{Headline scores}

\begin{figure}
\includegraphics[width=\textwidth]{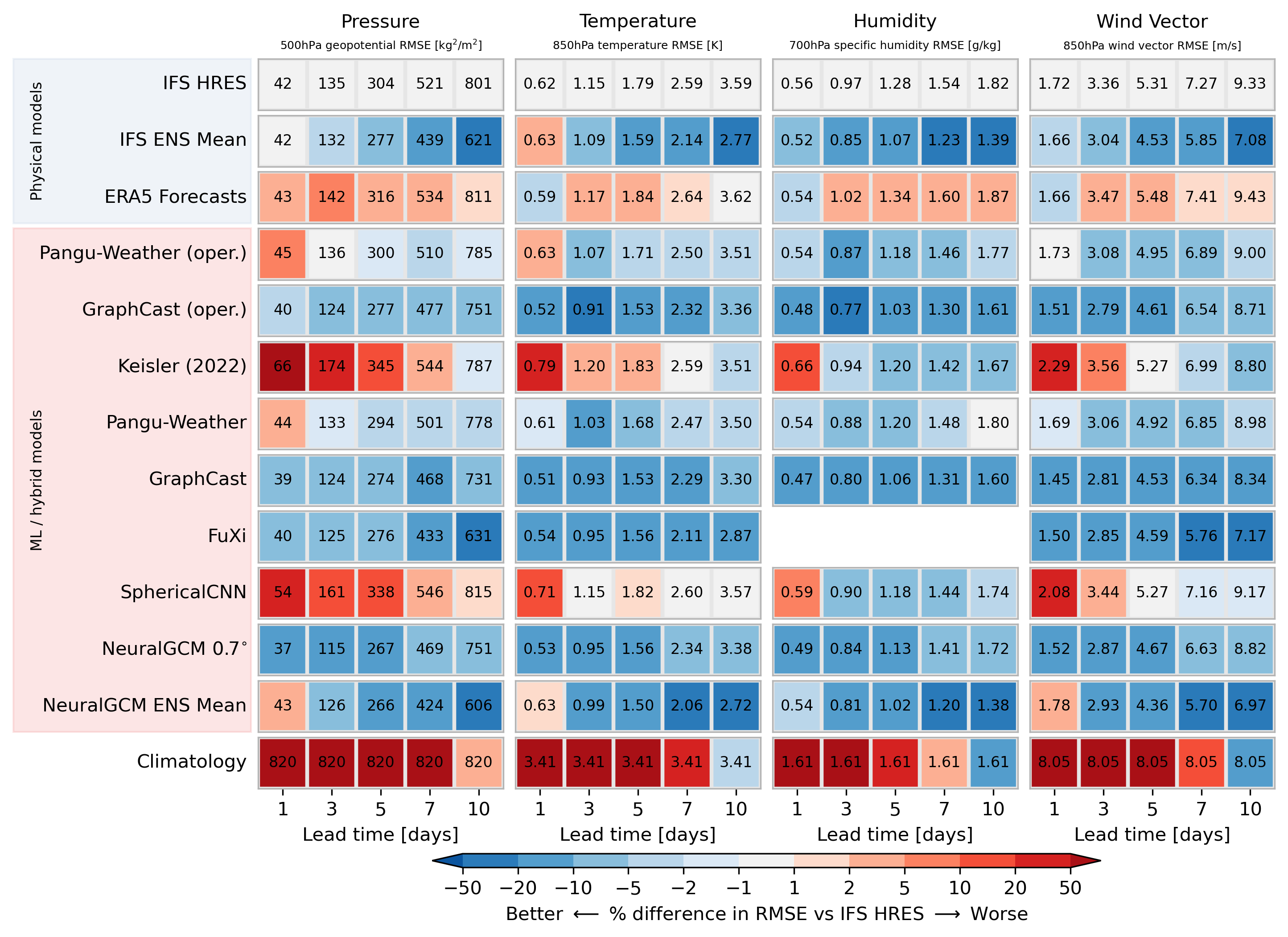}
\caption{Deterministic headline scorecards for upper-level variables. Values show absolute RMSE. Colors denote \% difference to the IFS HRES baseline.}
\label{fig:scorecard_upper}
\end{figure}

\begin{figure}
\includegraphics[width=\textwidth]{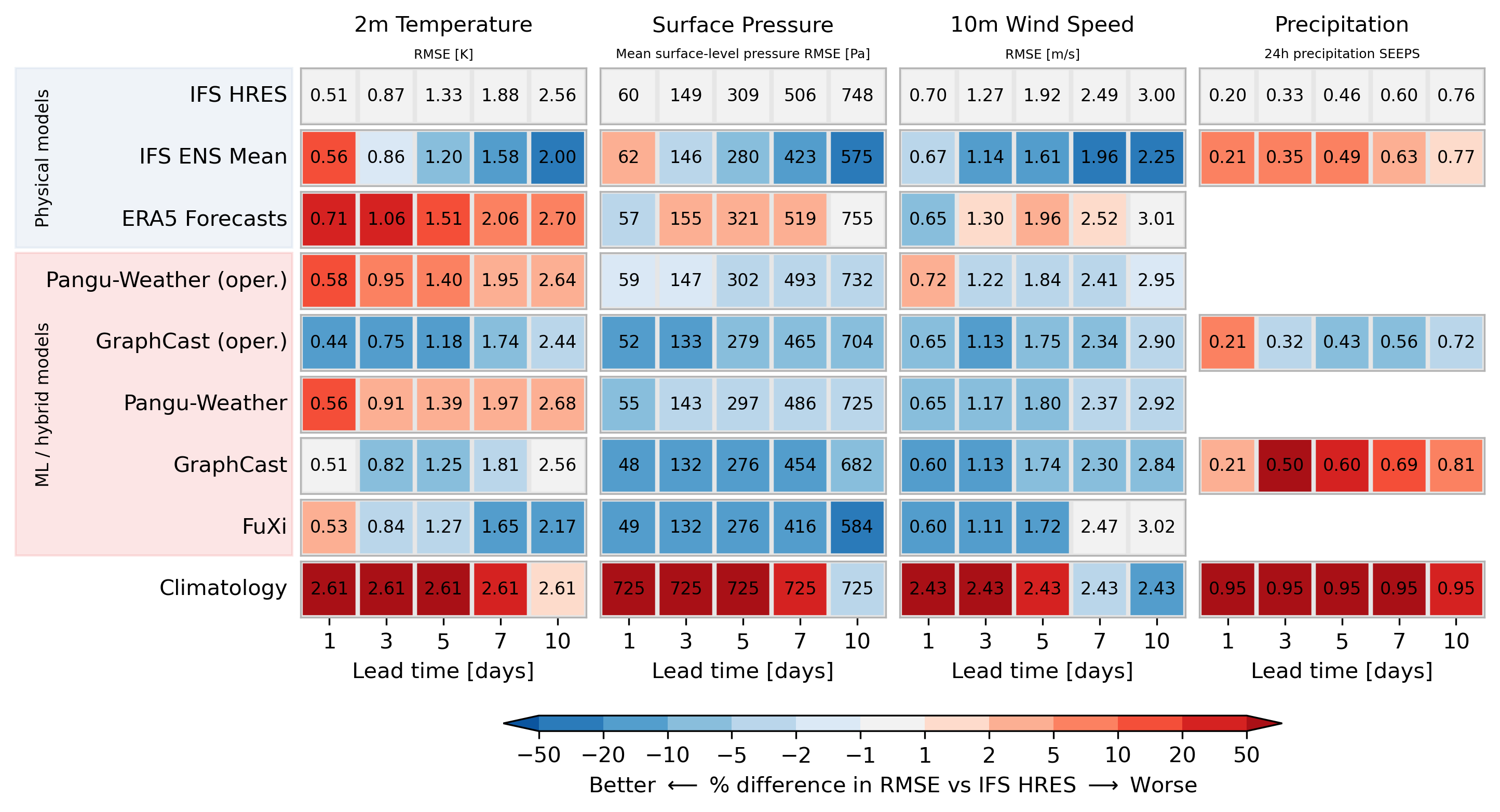}
\caption{Deterministic headline scorecards for surface variables. Values show absolute RMSE, with the exception of precipitation which shows SEEPS (evaluated against ERA5 in all cases). Colors denote \% difference to the IFS HRES baseline.}
\label{fig:scorecard_surface}
\end{figure}

\begin{figure}
\includegraphics[width=\textwidth]{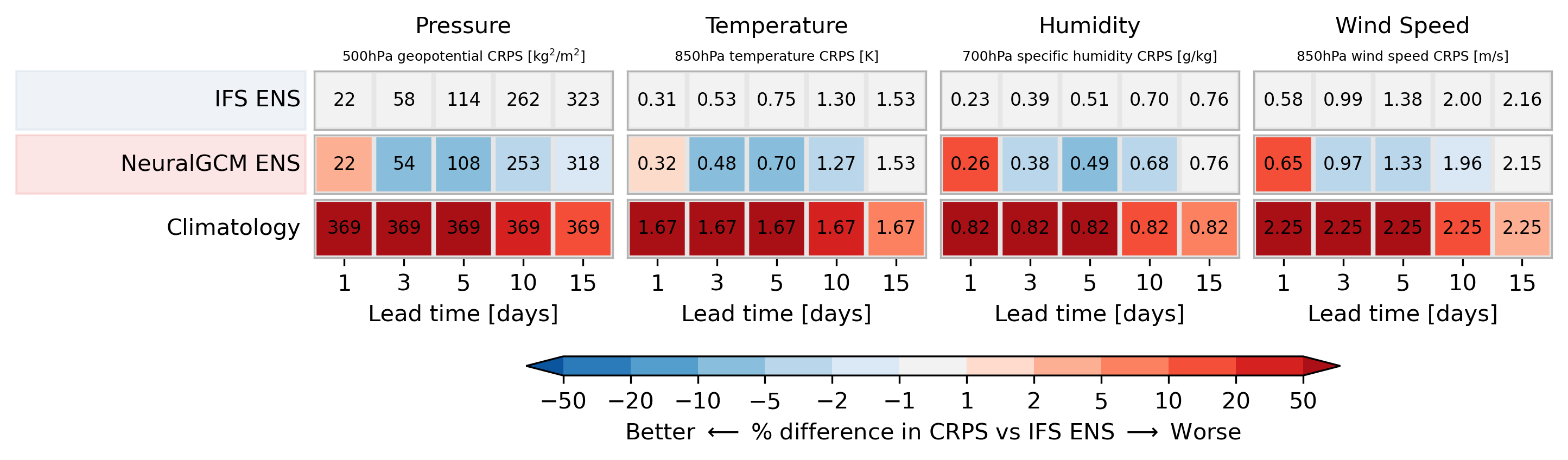}
\caption{Probabilistic headline scorecards for upper-level variables. Values show absolute CRPS. Colors denote \% difference to the IFS ENS baseline.}
\label{fig:scorecard_prob}
\end{figure}

\begin{figure}
\includegraphics[width=\textwidth]{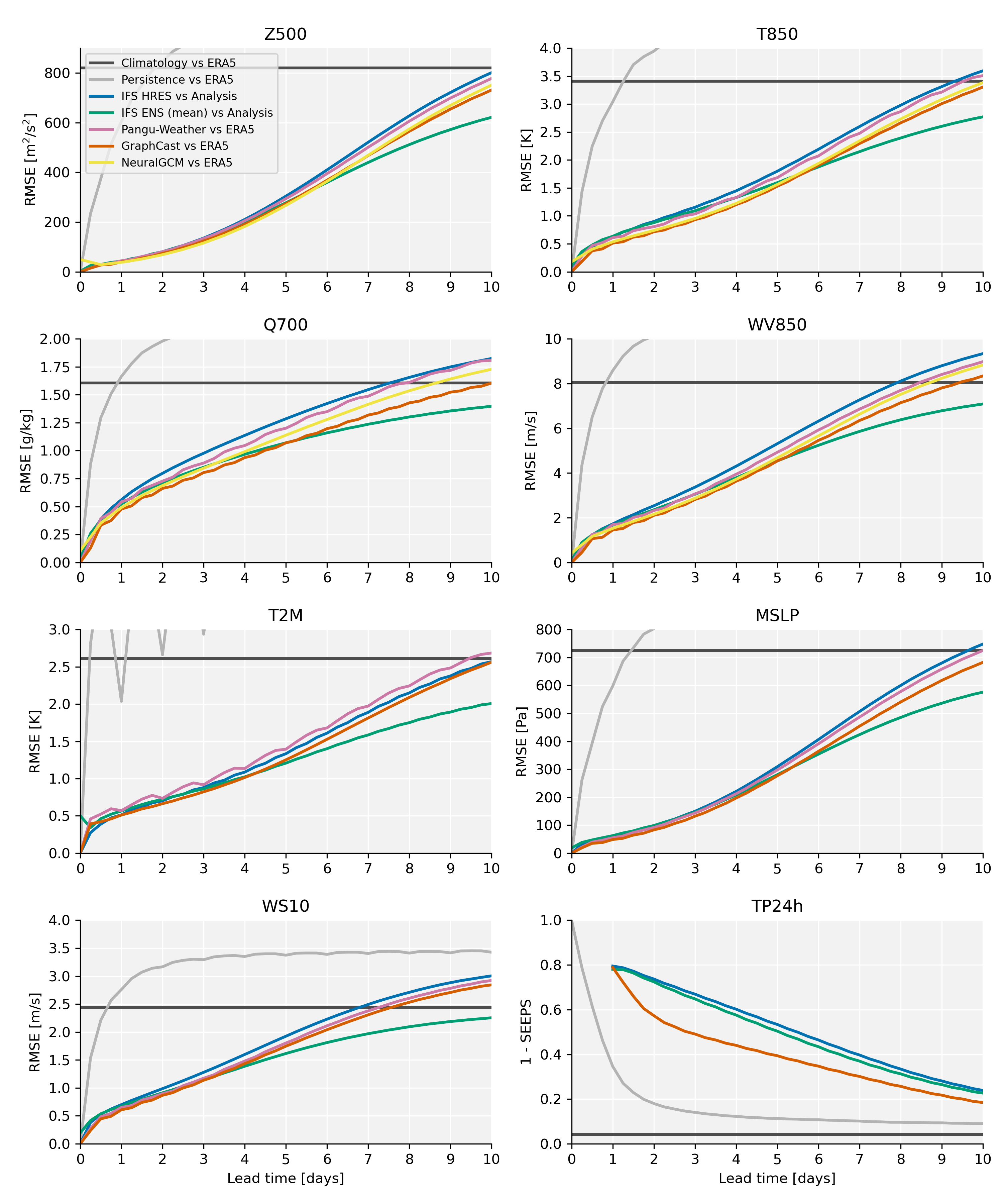}
\caption{Global RMSE (SEEPS for TP24h) for headline variables for the year 2020. Note that for TP24h, IFS HRES and IFS ENS (mean) are evaluated against ERA5, since no precipitation accumulations are available for the analysis. Not all models/datasets have all variables available.}
\label{fig:rmse_abs}
\end{figure}

\begin{figure}
\includegraphics[width=\textwidth]{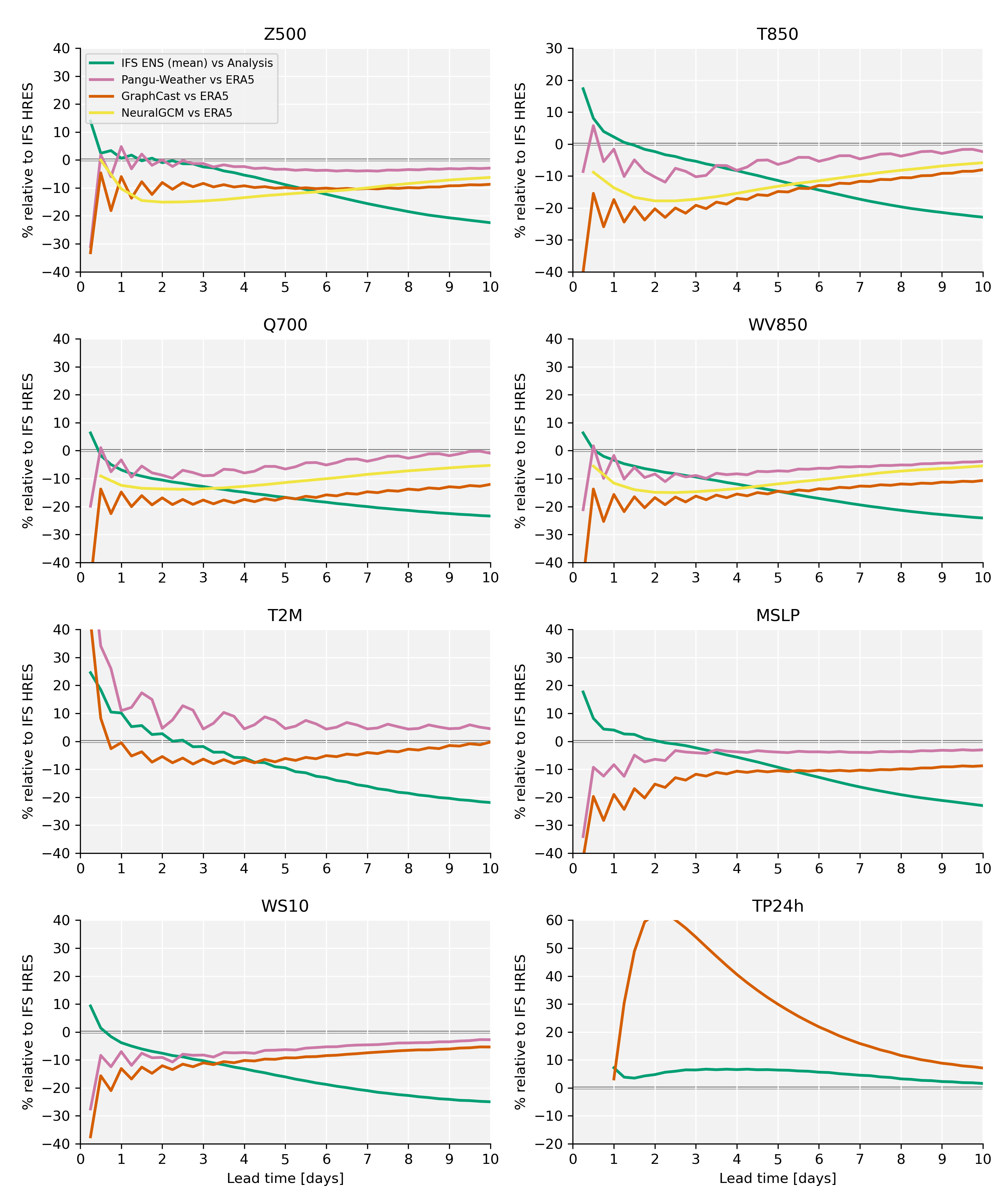}
\caption{Global RMSE/SEEPS \% difference compared to IFS HRES for headline variables for the year 2020. Negative values indicated lower RMSE. Note that for TP24h, IFS HRES and IFS ENS (mean) are evaluated against ERA5, since no precipitation accumulations are available for the analysis, and the metric is SEEPS (in this case not 1-SEEPS to have a consistent orientation with the other relative plots). Not all models/datasets have all variables available.}
\label{fig:rmse_rel}
\end{figure}

\begin{figure}
\includegraphics[width=\textwidth]{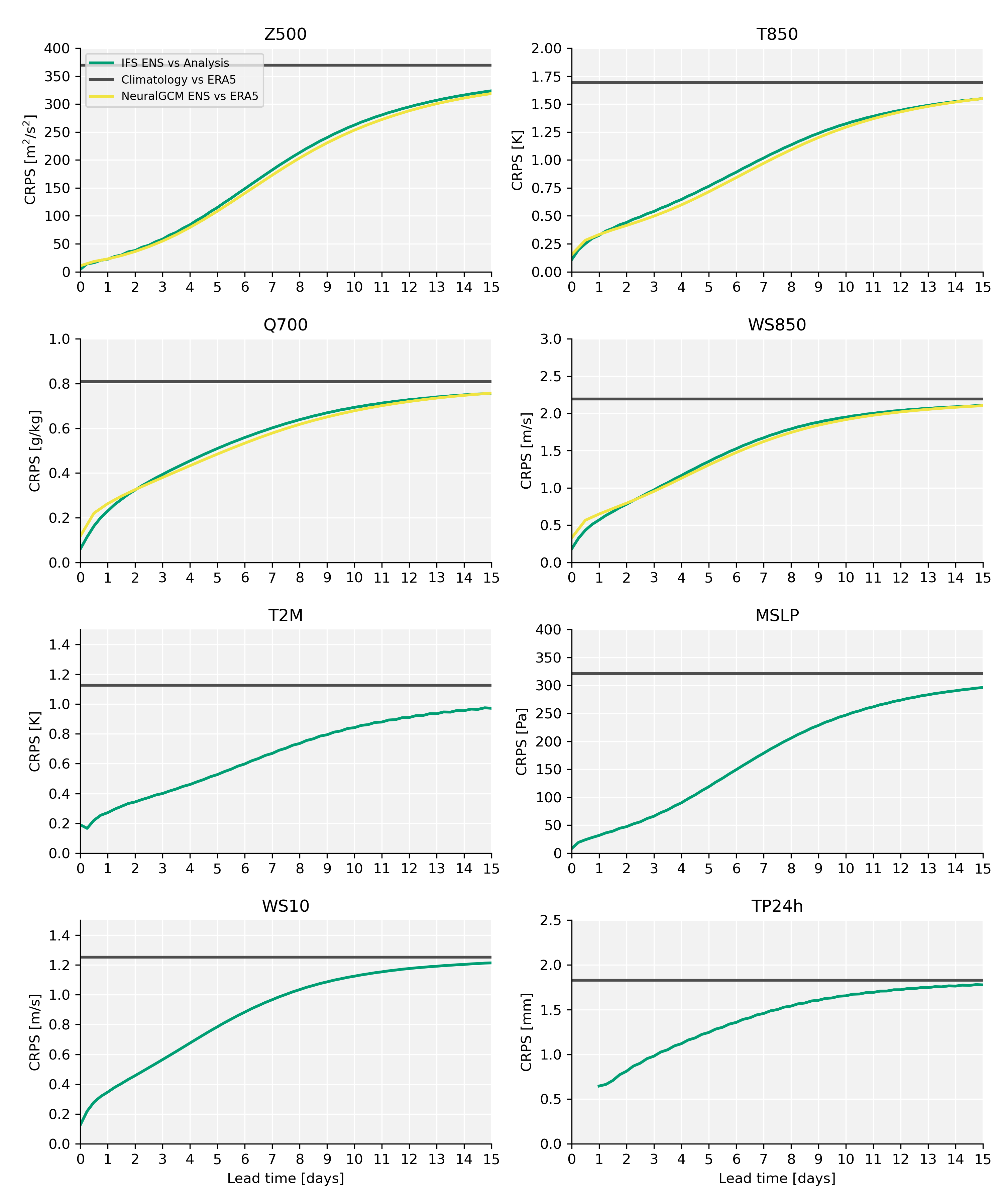}
\caption{Global CRPS for headline variables for the year 2020.}
\label{fig:crps_abs}
\end{figure}

Figs.~\ref{fig:scorecard_upper} and \ref{fig:scorecard_surface} show scorecards of the deterministic headline scores relative to the IFS HRES baseline. Fig.~\ref{fig:scorecard_prob} shows a scorecard of upper-level headline scores. Figs.~\ref{fig:rmse_abs}, \ref{fig:rmse_rel} and \ref{fig:crps_abs} show the deterministic and probabilistic scores as a function of lead time, respectively. Scores for the other metrics, variables and resolutions can be found on the WeatherBench 2 website.

On RMSE, IFS ENS (mean) performs better than IFS HRES for longer lead times. For smaller scale variables this transition happens earlier. For example, for larger scale variables like pressure and temperature the IFS ENS (mean) has lower errors up to around 2 days, compared to up to around 12h for wind and humidity. The state-of-the-art, deterministic, data-driven methods (Pangu-Weather, GraphCast, FuXi and NeuralGCM) also score well compared to IFS HRES, with scores similar to those shown in the respective papers. In comparison to IFS ENS (mean), these deterministic models have lower errors up to 3--6 days, after which the ensemble mean has the lowest error. NeuralGCM ENS (mean) roughly matches the IFS ENS (mean) scores on RMSE.

It is also noticeable, especially in the plots relative to IFS HRES, that the scores for the data-driven methods have a strong 6h zig-zag pattern. This is an artifact of training and evaluating with ERA5, which has a 12h assimilation window. Forecasts initialized at 00/12UTC are initialized towards the beginning of the 09--21/21--09UTC ERA5 assimilation windows. This means that for the first 6h forecast, e.g., from 00 to 06 UTC, all the AI models have to learn is to emulate the IFS model version used in ERA5. For the next 6h window, e.g., from 06 to 12UTC, the models also have to learn the assimilation step---a harder task. If forecasts initialized at 06/18UTC were evaluated, the zig-zag pattern would be reversed. Note that in \cite{lam_learning_2023}, this pattern is not visible because evaluation is done on a 12h interval, thereby only sampling every second point shown here. Here we chose to show the 6h evaluation for completeness (except for NeuralGCM where we only had 12-hourly data).

Note that the larger error of IFS ENS (mean) at time $t=0$ for 2m temperature stems from a difference in resolution between the ensemble (0.2$^{\circ}$) and the analysis (0.1$^{\circ}$). Results for the newly upgraded higher resolution ensemble (not included in WB2 yet) do not show this large initial error.

The precipitation verification using the SEEPS score shows that IFS HRES is the most skillful model with the ``blurrier'' models, such as  IFS ENS (mean), being less skillful. 
This shows the advantage of using a categorical score compared to an average score such as RMSE. On RMSE, the order of the models is reversed (not shown).

For the probabilistic models, IFS ENS and NeuralGCM ENS show very similar scores on upper-level variables. Their skill approaches that of the probabilistic climatology towards the end of the two week forecast horizon. This is in line with the expected $\sim$15 day limit of predictability of synoptic weather \citep{selz_estimating_2019}. For some variables, while the IFS ENS and NeuralGCM ENS skill curves flatten off, there is still a gap to the climatological skill. This could have several reasons: first, our methodology for computing the probabilistic climatology might not be a perfect representation of the true climatology; second; the forecast might still have some skill there. Especially for large-scale variables such as pressure and temperature, there is evidence of skill in the sub-seasonal range; and third, especially for temperature variables, climate change has a non-negligible effect over the 30 years used to compute the climatology.

\begin{figure}
\includegraphics[width=\textwidth]{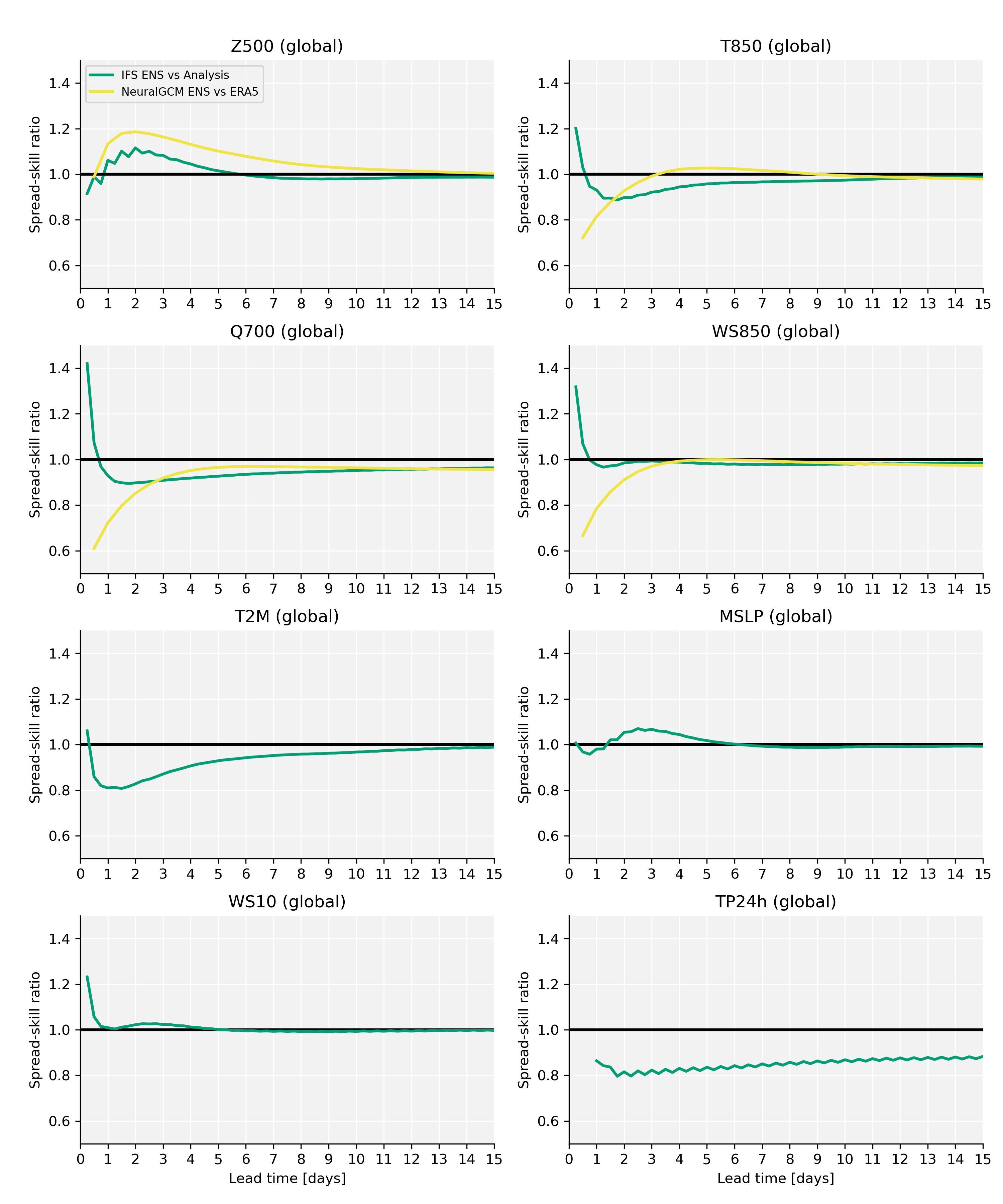}
\caption{Spread/skill ratio for headline variables for the year 2020.}
\label{fig:spread_skill_ratio}
\end{figure}

An analysis of the spread-skill relationship (Fig.~\ref{fig:spread_skill_ratio}) shows that except for precipitation, which is underdispersive, the IFS ensemble forecast is very well calibrated, especially for longer lead times. The initial hours show some spin-up with more spread in the first hours, followed by a ~2 day dip in the spread-skill ratio. For geopotential and pressure, this trend is reversed. NeuralGCM ENS also shows good calibration for longer lead times but instead appears to be underdispersive at early lead times, perhaps due to the lack of initial condition spread. Here it is important to not over-interpret global spread-skill ratios. More fine-grained analysis of tropical vs extra-tropical dispersion show more differentiated behavior. These results are included on the website.

\begin{figure}
\includegraphics[width=\textwidth]{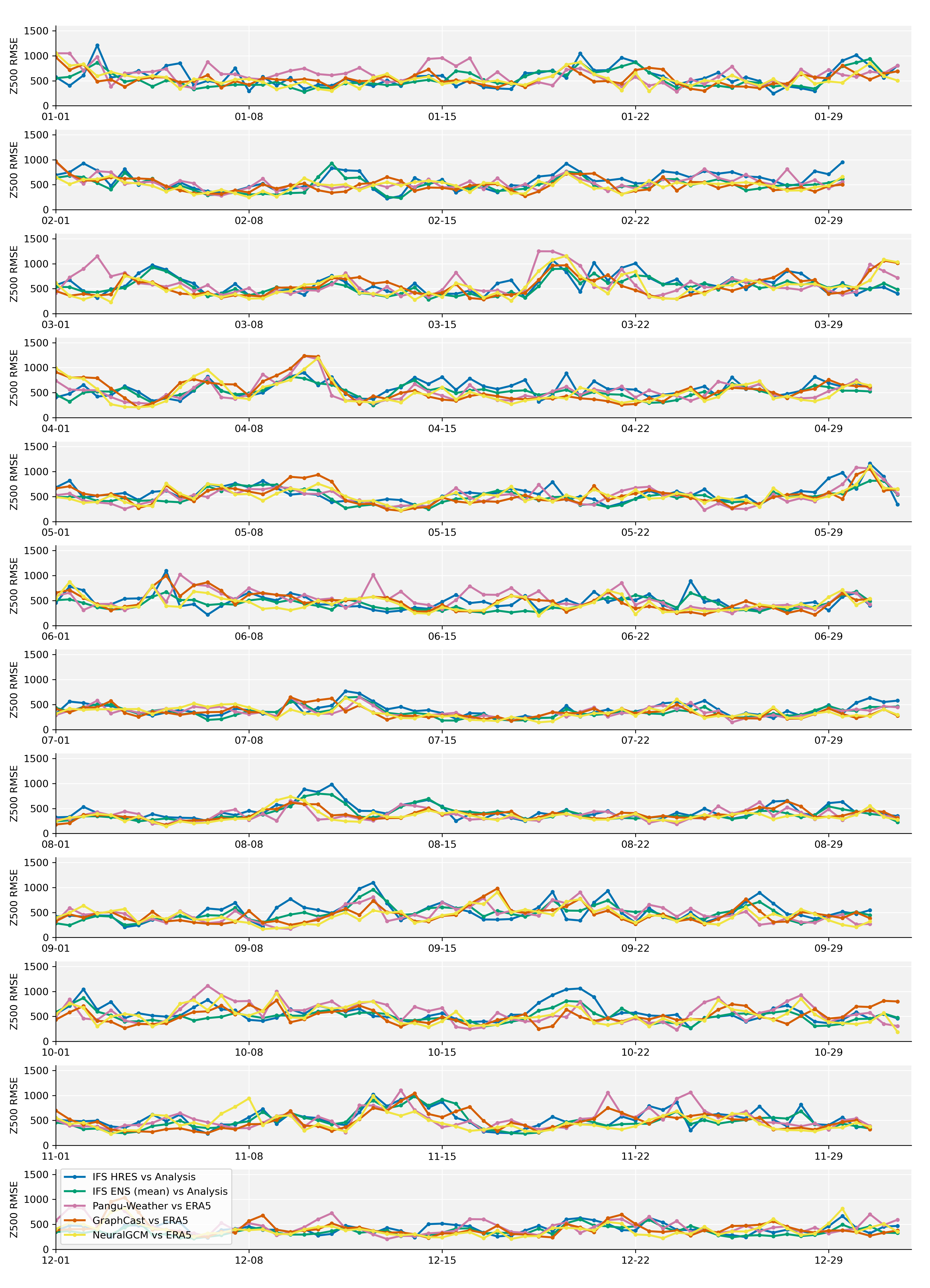}
\caption{Timeseries of day 6 RMSE in m$^2$/s$^2$ over Europe defined by $35^{\circ} < lat < 75^{\circ}$ and $-12.5^{\circ} < lon < 42.5^{\circ}$. All models evaluated at 1.5$^{\circ}$ resolution. Climatological and persistence errors are omitted.}
\label{fig:temporal_rmse}
\end{figure}

Fig.~\ref{fig:temporal_rmse} also shows a time series of 6 day RMSE over Europe for all of 2020 for each of the models. Here, several forecast bust events are visible, where forecast skill drops significantly below normal \citep{rodwell_characteristics_2013}. Interestingly, some bust cases are shared between the physical and AI models (e.g. around March 19) while for others AI models continue to perform well, despite IFS HRES and ENS performing badly (e.g. August 11). In some other cases, AI models spuriously perform badly. Generally, this implies that AI models show similar error characteristics to physical models.

\subsection{Bias}

\begin{figure}
\includegraphics[width=\textwidth]{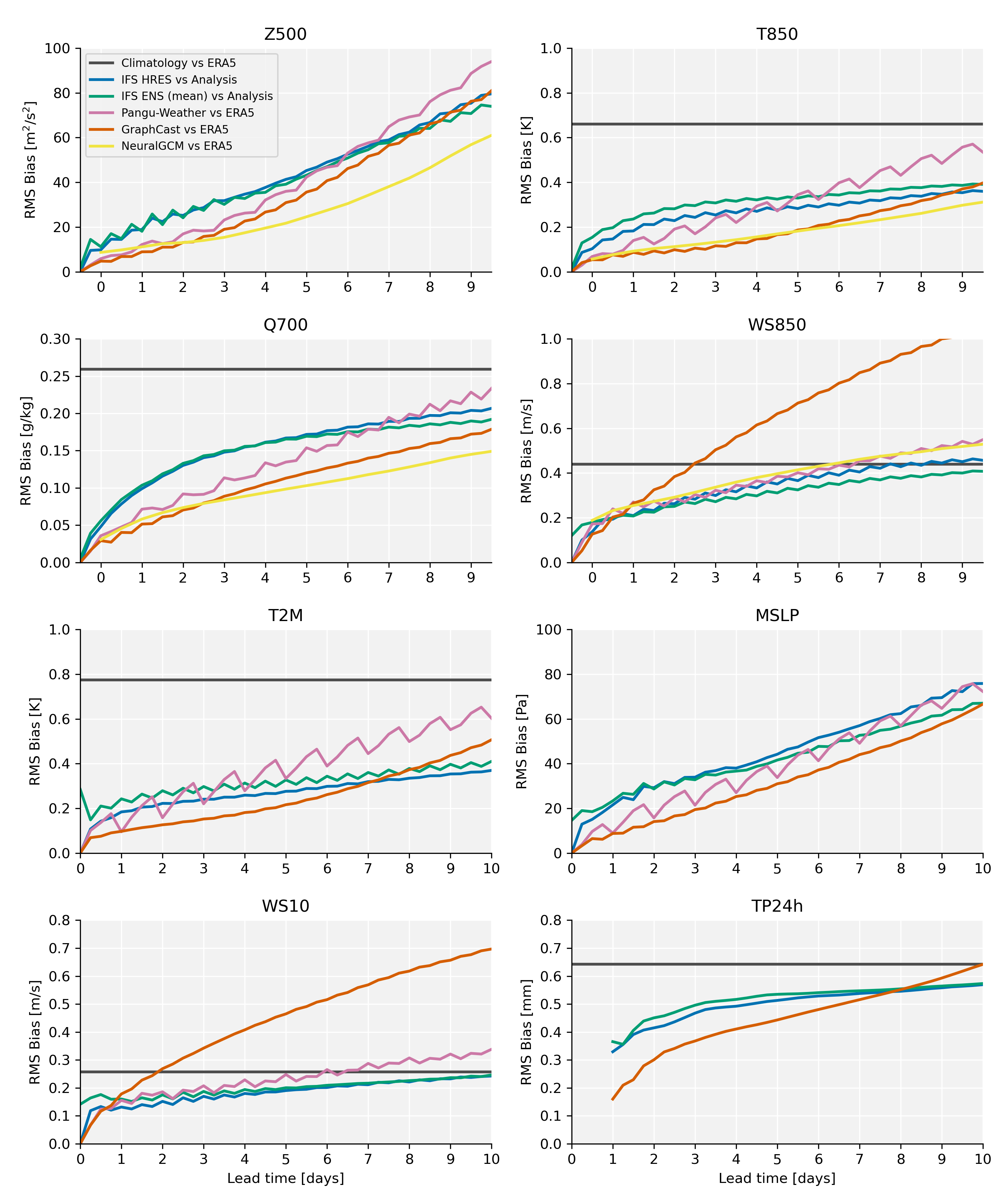}
\caption{Global RMSB for the headline variables for the year 2020. For Z500 and MSLP the climatological bias is comparatively large and is not shown within the figure limits}
\label{fig:rmsb}
\end{figure}

Fig.~\ref{fig:rmsb} shows the RMS bias (RMSB) for the headline variables. Additionally, bias maps for 2m temperature, 24h precipitation, 10m wind speed and 700hPa specific humidity can be found in the supplement. It is important to note that one year is a small sample to compute bias statistics. In the bias maps in the supplement it is evident that dominant weather patterns in 2020 influence the bias. For example, 2020 saw a persistent heat wave in Siberia. All models tend to have a cold bias there since they fail to fully predict such a persistent pattern. Therefore, not too much should be interpreted into regional biases without considering the context. 

What is noticeable is that while ML methods tend to have a lower average bias compared to physical methods, especially at early lead times, for wind speed GraphCast (and Keisler (2022); not shown) show a large increase in average bias with lead time. The bias maps (Figs.~S6 and S7) show that this is due to a consistent bias towards smaller wind speed values. This bias is not present when looking at U or V separately (see supplement of \cite{lam_learning_2023}). What this indicates is that ML models trained to minimize each wind component separately, and doing a good job at doing so, can struggle to represent correlations between those components. Interestingly, Pangu-Weather does not have this bias. Neither does the hybrid NeuralGCM model.

\subsection{Spectra}

\begin{figure}
\includegraphics[width=\textwidth]{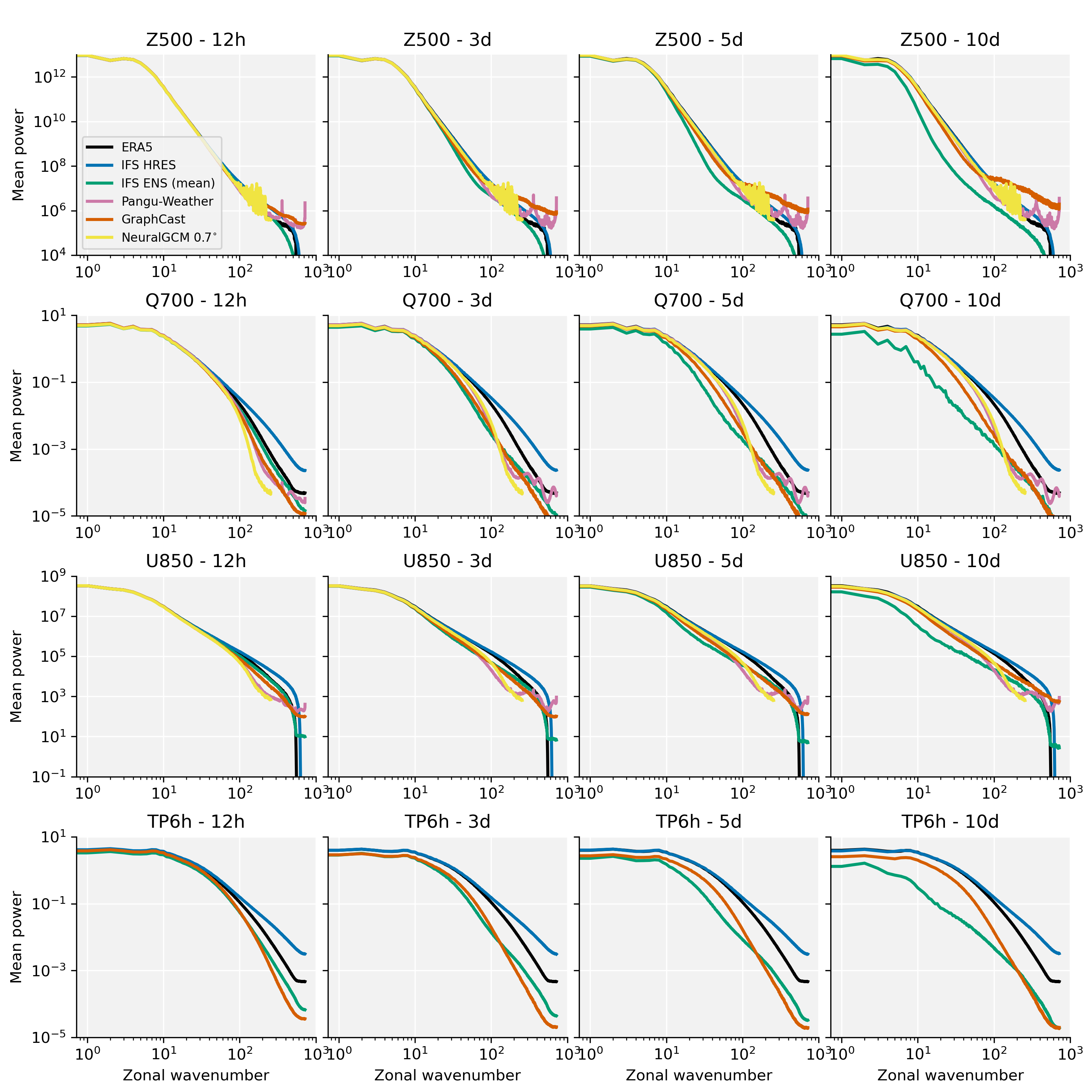}
\caption{Power spectra of 500hPa geopotential, 700hPa specific humidity, 850hPa u-wind and 6h precipitation accumulation for different lead times.}
\label{fig:spectra}
\end{figure}

Fig.~\ref{fig:spectra} shows the power spectra for four variables and four lead times. The ERA5 spectrum can be taken as a reference to compare between lead times as it is constant with lead time. IFS HRES generally has more energy on small scales compared to ERA5, likely due to its higher resolution, but is also quasi-constant with lead time which is to be expected for a physical model. The IFS ENS (mean) shows an increasing drop-off in power with lead time, starting with the smaller wavelengths and then also for the larger wavelengths for longer lead times. Here it is important to re-iterate that the ensemble mean does not represent a realistic realization of weather but rather the mean of the forecast distribution. In other words, the mean will become increasingly smooth with lead time, eventually converging to the climatological mean. The progression of the drop-off from smaller towards longer wavelengths is consistent with the model of upscale error growth first proposed by \cite{lorenz_predictability_1969}. For variables that vary more on small scales, such as humidity and precipitation, this effect is larger.

For Z500, the AI models tend to have more energy on small scales, more so for longer lead times. This increase in small scale energy for Z500 is in contrast to the behavior for the other variables where the AI models show a significant drop in small scale variability from 6h to 3d. This is a result of the aforementioned smoothing. After 3 days most AI models tend to have a constant spectrum. This is in line with the 24--72 h optimization windows. In other words, these models tend to optimize their blurring for those time ranges. The ensemble mean continues to become blurrier resulting in lower RMSE values for longer lead times. The spectra of NeuralGCM show less variance on small scales due to the model's lower resolution but the spectra are roughly consistent with lead time.

The spectra clearly show some of the blurring exhibited by some AI models. It is important to note though that having a spectrum that matches observations well is a necessary but not sufficient conditions for ``realism''.

\subsection{Case studies}

\begin{figure}
\includegraphics[width=\textwidth]{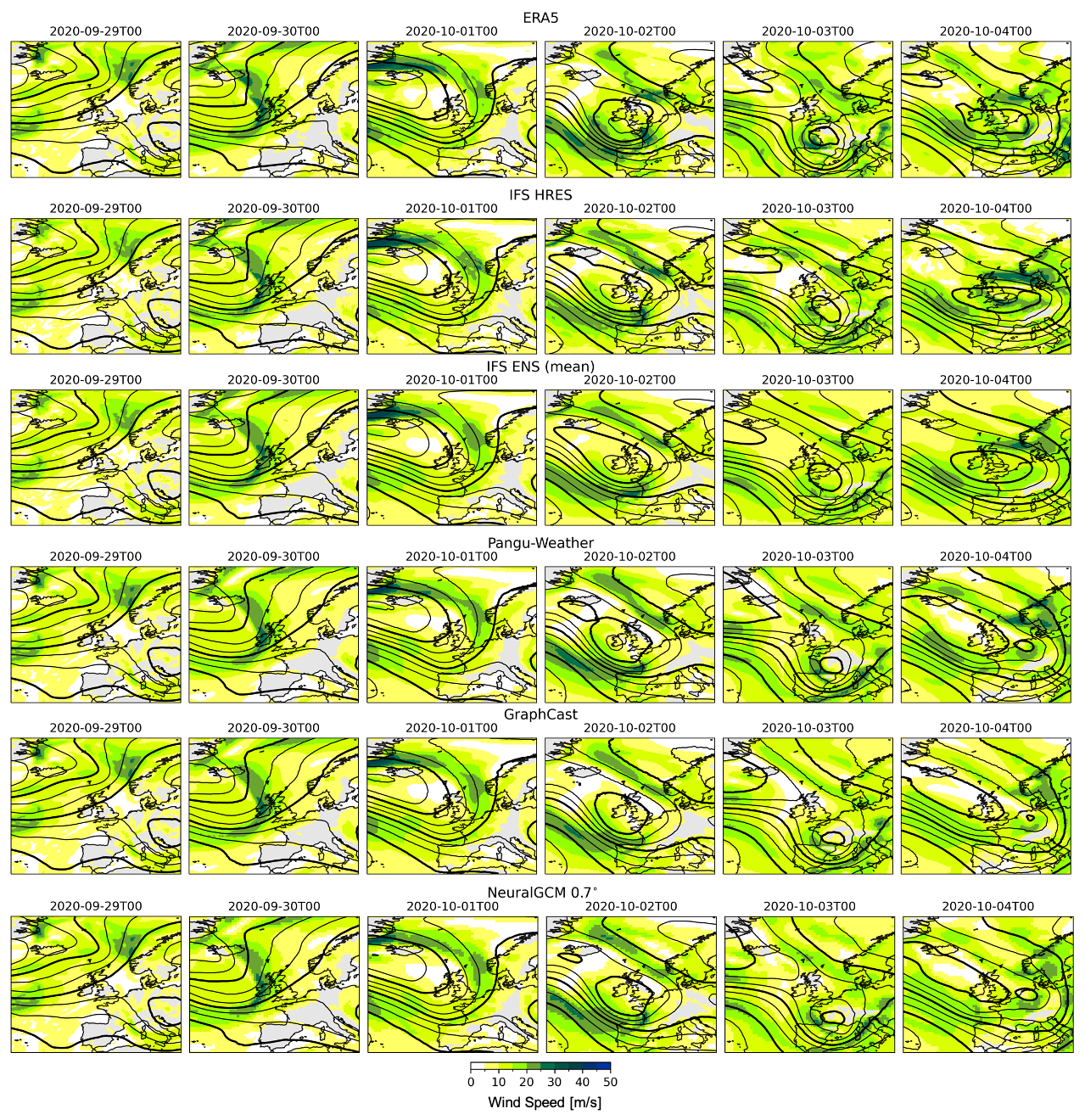}
\caption{Case study: Storm Alex. Top row shows ERA5 ground truth. Other rows show forecasts initialized at 2020-09-29 00UTC. Lines show Z500 contours. Shading shows 850hPa wind speed.}
\label{fig:case_alex}
\end{figure}

While case studies can only provide anecdotal evidence of model performance, they offer a more ``human'' and holistic view that skill scores cannot provide. Here we shows one such case study. Another case study of Hurricane Laura can be seen in Fig.~S10 On the WB2 website, we eventually plan to extend this catalog of case studies. 

Fig.~\ref{fig:case_alex} shows model forecasts and the corresponding ERA5 ground truth for storm Alex which brought damaging wind and rain across Europe (\url{https://climate.copernicus.eu/esotc/2020/storm-alex}). What is noticeable is that all models, physical and AI-based, predict the evolution of the storm with impressive accuracy and agreement up to four days lead time. October 2 saw strong winds over the North-West of France as well as parts of northern Spain, which was predicted well by all methods. On October 4, a cut-off trough developed cetered over the English channel. IFS HRES and even the ENS mean reasonably predict this whereas the AI methods favor a continuous trough stretching from Iceland to England.

Fig.~S10 shows the evolution of Hurricane Laura, the most damaging hurricane of the 2020 season (\url{https://www.nhc.noaa.gov/data/tcr/AL132020_Laura.pdf}). IFS HRES shows impressive skill in predicting the evolution and track of the hurricane up to 5 days in advance with the timing and location of landfall on August 8 almost perfectly forecasted.  The IFS ENS mean, somewhat predictably, fails to predict the strength and location of the hurricane because it averages out several hurricane tracks. Pangu-Weather and GraphCast have a solid track forecast with reasonable cyclone structure but fail to predict the intensity in pressure and wind speed seen in the actual hurricane of the IFS forecast. GraphCast predicts landfall West of the actual location. NeuralGCM, for this case, shows a remarkable agreement with ERA5, on par with IFS HRES.

We plan to add more case studies to the website in the future. While case studies should not be over-interpreted, they are evidence that while sometimes lacking in detail and fidelity, AI models can produce skillful extreme weather forecasts for events outside of their training dataset. 

\section{Discussion}
\label{sec:discussion}

\subsection{ERA5 as ground truth}
\label{sec:discussion_era5}
Here, we use ERA5 as our ground truth dataset. However, as already discussed in Section~\ref{sec:data_era5}, ERA5 is a model simulation that is kept close to observations, rather than direct observations. The quality of ERA5 depends on the variable in question. Large-scale dynamical variables like geopotential and upper-levels temperatures and wind tend to be well represented in reanalysis datasets. For surface variables, the sparsity of observations and difficulty of representing small-scale physics in the underlying model can cause larger discrepancies with observations. This is especially true for precipitation, which is not directly assimilated into ERA5 (e.g., through radar observations) and often show large differences to rain gauges or radar precipitation estimates (e.g., \citep{lavers_evaluation_2022}, \citep{andrychowicz_deep_2023}). The precipitation evaluation using ERA5 shown here should really be seen as a placeholder for more accurate precipitation data.

Operational weather services like ECMWF verify their forecasts with direct observations, e.g., from weather stations, in addition to using assimilated ground truths. This is something we are looking to add to WeatherBench in the future. However, station observations come with their own issues: first, they are unevenly distributed with some regions being especially sparse in observations; second, station data comes in varying quality and requires careful quality control; and third, station data suffers from representativeness issues; that is, the station might not be representative of the model grid box it is compared to. 

For these reasons, ERA5 still represents a good option for evaluation for most variables because of its temporal and spatial coverage but it should be kept in mind that it does not tell the full story. Notably, some regional ML and post-processing models already look at using data directly for training and evaluation \citep{andrychowicz_deep_2023, demaeyer_euppbench_2023}.

\subsection{ERA5 initialization versus operational forecasting}
\label{sec:discussion_ic}
Most data-driven methods so far have been trained and evaluated using ERA5 data as initial conditions. As already discussed in Section~\ref{sec:data_era5}, ERA5 data would not be available in real time to initialize live, operational forecasts. For a true apples-to-apples comparison with operational physical models, ML forecasts would need to be initialized using operationally available initial conditions. The first such experiment has been done by \cite{ben-bouallegue_rise_2023}, who took the Pangu-Weather model trained on ERA5 and initialized it with operational IFS analyses. In addition, we include operational versions of Pangu-Weather and GraphCast in the WeatherBench scorecards. Across the board, the scores are very similar between the ERA5-initialized and operational versions, whether they have been fine-tuned (in the case of GraphCast) or not (in the case of Pangu-Weather). This suggests that the distribution shift between ERA5 and the operational IFS analysis is manageable for these AI models. Regardless, going forward it will be important to distinguish between ERA5-initialized and (quasi-)operational models as e.g. done in Table~\ref{tab:models}. For an apples-to-apples comparison, we encourage researchers to run and evaluate their models with operational analyses.

\subsection{Forecast realism, probabilistic forecasts and extremes}
\label{sec:discussion_realism}
One feature of current AI methods is that they produce unrealistically smooth forecasts, a trend that often becomes more pronounced with lead time. This is a direct consequence of these models minimizing a deterministic mean error (such as MSE). Because the evolution of the atmosphere is chaotic and forecast uncertainty grows with time, this will lead models to ``hedge their bets'' by predicting the mean of the potential forecast distribution. In a non-linear, high-dimensional system such as the atmosphere, the mean of many states is not a realistic state itself. The result are blurry forecasts that perform well on some skill metrics (like RMSE or ACC) but do not represent a possible weather state. This is evident in several analyses presented here: spectra show excessive blurring of AI models for longer lead times; the SEEPS score shows how blurry models fail to predict the right precipitation category; wind speed biases are evidence that current AI models have difficulties learning correlations between variables; and the case studies show that the intensity of local features such as wind speed, precipitation and cyclone intensity aren't represented. For some applications, having good forecasts of average weather is sufficient but for others AI models are not yet appropriate. 

This is naturally related to probabilistic prediction, i.e., predicting the full range of potential weather outcomes, which is so important for predicting the probability of extreme weather. Current, ``traditional'' forecast systems use ensembles for this purpose. AI methods could follow a similar approach by producing an ensemble of generative roll-outs, or they could directly predict probabilistic outcomes of interest, such as the probability distribution of precipitation as done in the MetNet family of models \citep{andrychowicz_deep_2023} or many post-processing models (e.g. \cite{gneiting_calibrated_2005}). For many applications, such as predicting weather at a particular location, the latter approach might be more straightforward and sufficient, while for other applications, for example cyclone track forecasting or when humans interpret model output, temporal and spatial structure is important. Working closely with end users will be key in determining the most appropriate probabilistic representation for each application.

The probabilistic evaluation metrics proposed in WB2 (CRPS and spread/skill) are very much just the tip of the iceberg. They are univariate statistics, i.e., they ignore spatial and temporal correlations, and they do not specifically focus on extreme weather. Just like with the deterministic metrics (RMSE or ACC), better CRPS values do not automatically mean a more useful forecast. Weather forecasting is a very high-dimensional problem, which means that there won't be a single metric to determine forecast quality. Rather, evaluation will have to be guided by applications. 

\subsection{Post-processing}
In this paper, the focus has been on data-driven forecast models. However, AI can also and has for a long time been used to post-process dynamical forecasts \citep{rasp_neural_2018, finn_self-attentive_2021, gneiting_strictly_2007, gronquist_deep_2021}. WB2 can equally well serve as a benchmark for post-processing models, on top of dynamical forecasts such as those produced by ECMWF. In fact, comparing ``purely'' data-driven forecasting models with dynamical models with state-of-the-art post-processing should be an insightful exercise going forward. Note that benchmarks for post-processing have been proposed by \cite{ashkboos_ens-10_2022} including re-forecasts for training post-processing models and an extreme weighted CRPS score definition. Similarly, \cite{demaeyer_euppbench_2023} present a benchmark for station-based post-processing.

Significant improvements over raw dynamical model output can be expected using post-processing. Previous studies \citep{rasp_neural_2018, ben-bouallegue_improving_2023, finn_self-attentive_2021} suggest that probabilistic forecasts can be improved by up to 20\% in terms of CRPS, depending on the variable in question. Since data-driven methods, at least partly, already perform a post-processing implicitly, a fair comparison would be against post-processed dynamical models. Another interesting question is how much AI models can benefit from additional post-processing. Hopefully, these comparisons will be added to WB2 soon.

\section{Conclusion}
WeatherBench 2.0 is an updated benchmark for data-driven, global weather forecasting. It is motivated by the rapid advances in used AI methods since the publication of the original WeatherBench benchmark. WB2 is designed to stay close to the operational forecast evaluation used by many weather centers and to provide a robust framework for evaluating new methods against operational baselines. By providing evaluation code and data, we hope to speed up machine learning workflows and ensure the reproducibility of results. 

WB2 is also designed to be a dynamic framework that will be updated with new metrics and models as this area of research evolves. Several possible extensions have already been discussed in this paper, for example including station observations and evaluating extremes.

\newpage
\small
\bibliography{references_20240110.bib}
\bibliographystyle{apalike}

\newpage
\section*{Supplement}
\setcounter{figure}{0}
\renewcommand{\thefigure}{S\arabic{figure}}
\setcounter{section}{0}
\renewcommand{\thesection}{S\arabic{section}}

\renewcommand{\floatpagefraction}{0.0}
\begin{figure}
\includegraphics[width=\textwidth]{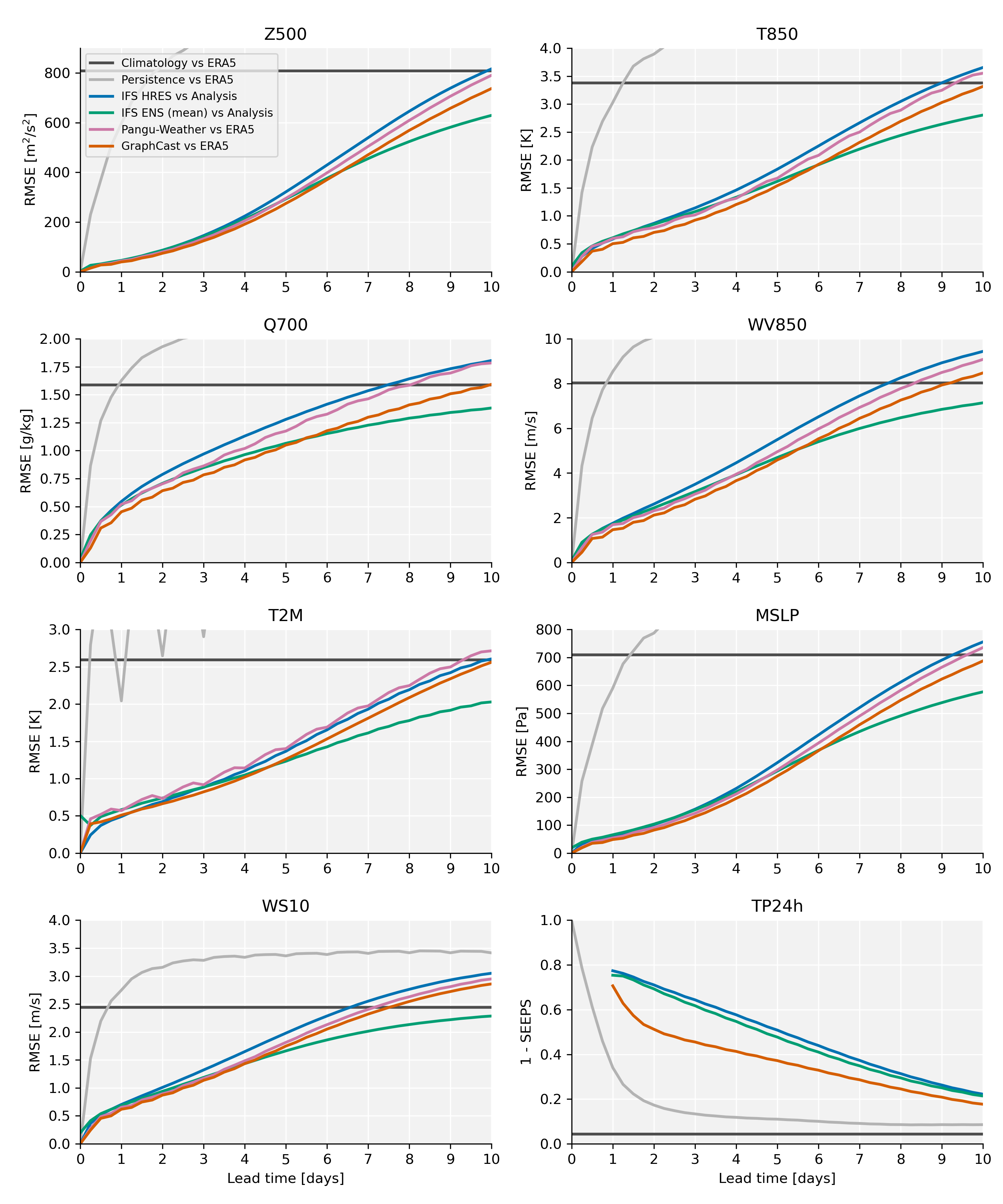}
\caption{Global RMSE/SEEPS values for the year 2018.}
\label{fig:acc_abs}
\end{figure}





\begin{figure}
\includegraphics[width=\textwidth]{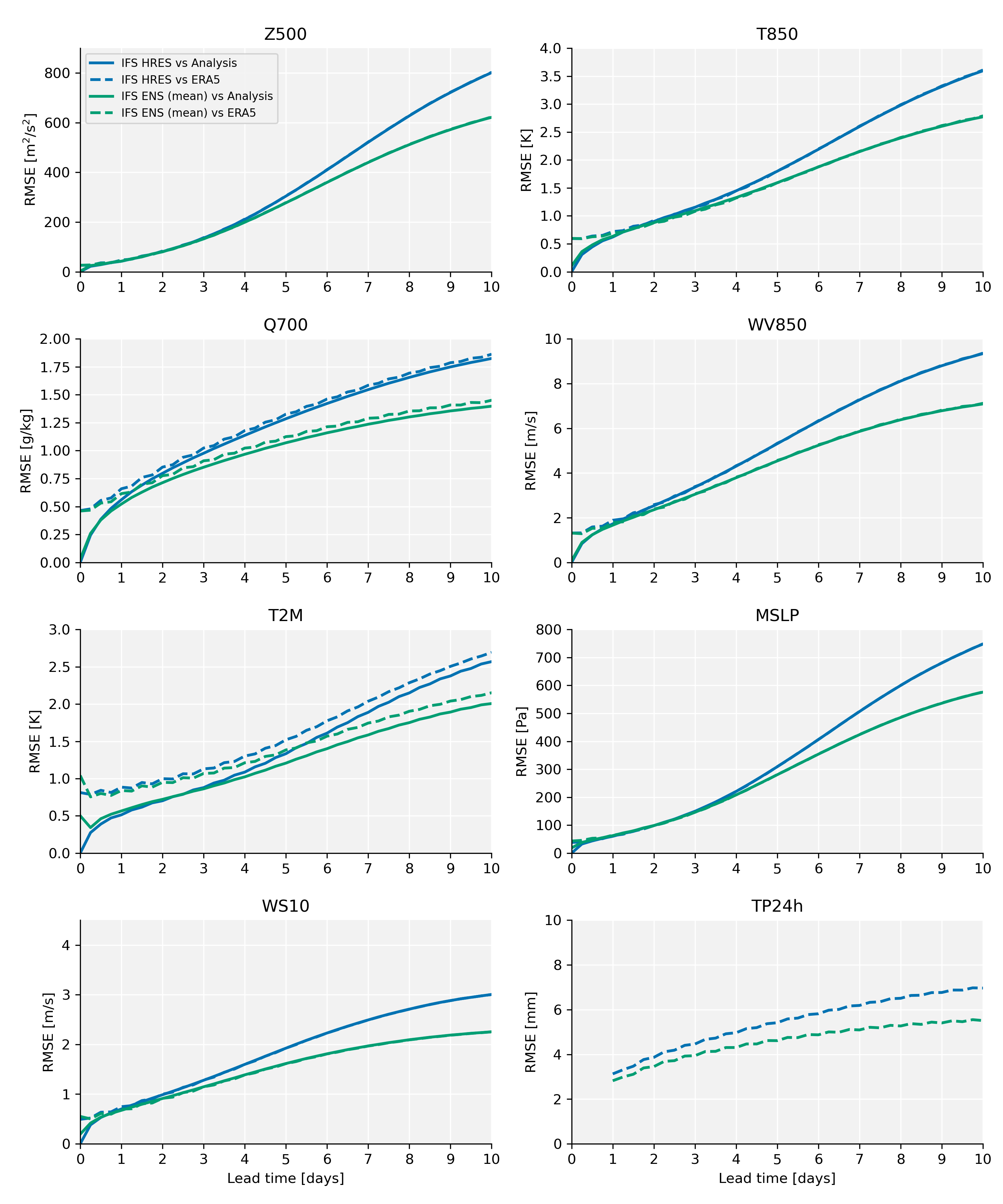}
\caption{Comparison of evaluating IFS HRES and ENS (mean) with analysis and ERA5 ground truth.}
\label{fig:analysis_vs_era}
\end{figure}
\newpage

\begin{figure}
\includegraphics[width=\textwidth]{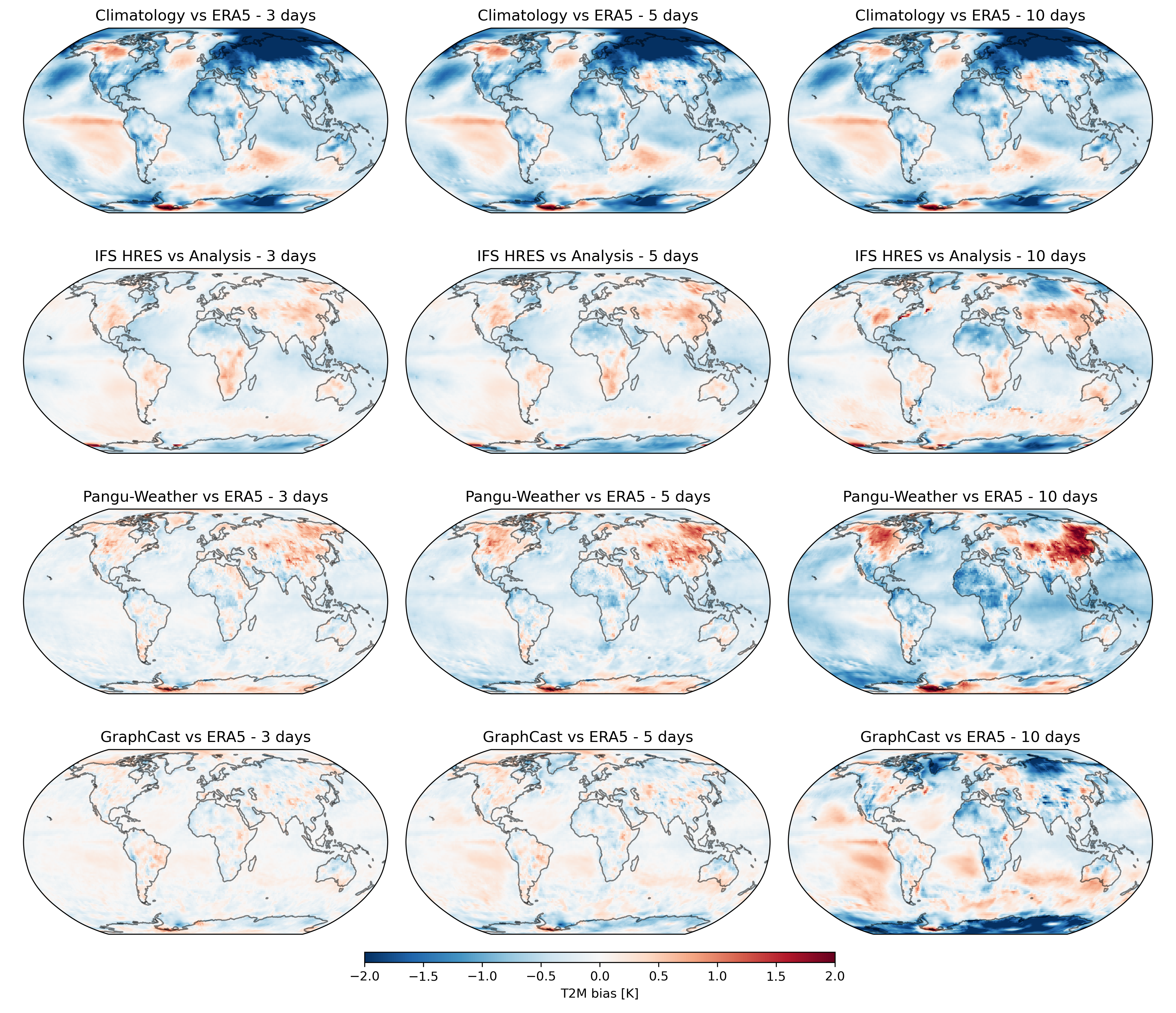}
\caption{Global mean 2m temperature bias for 3, 5 and 10 day lead times.}
\label{fig:bias_t2m}
\end{figure}

\begin{figure}
\includegraphics[width=\textwidth]{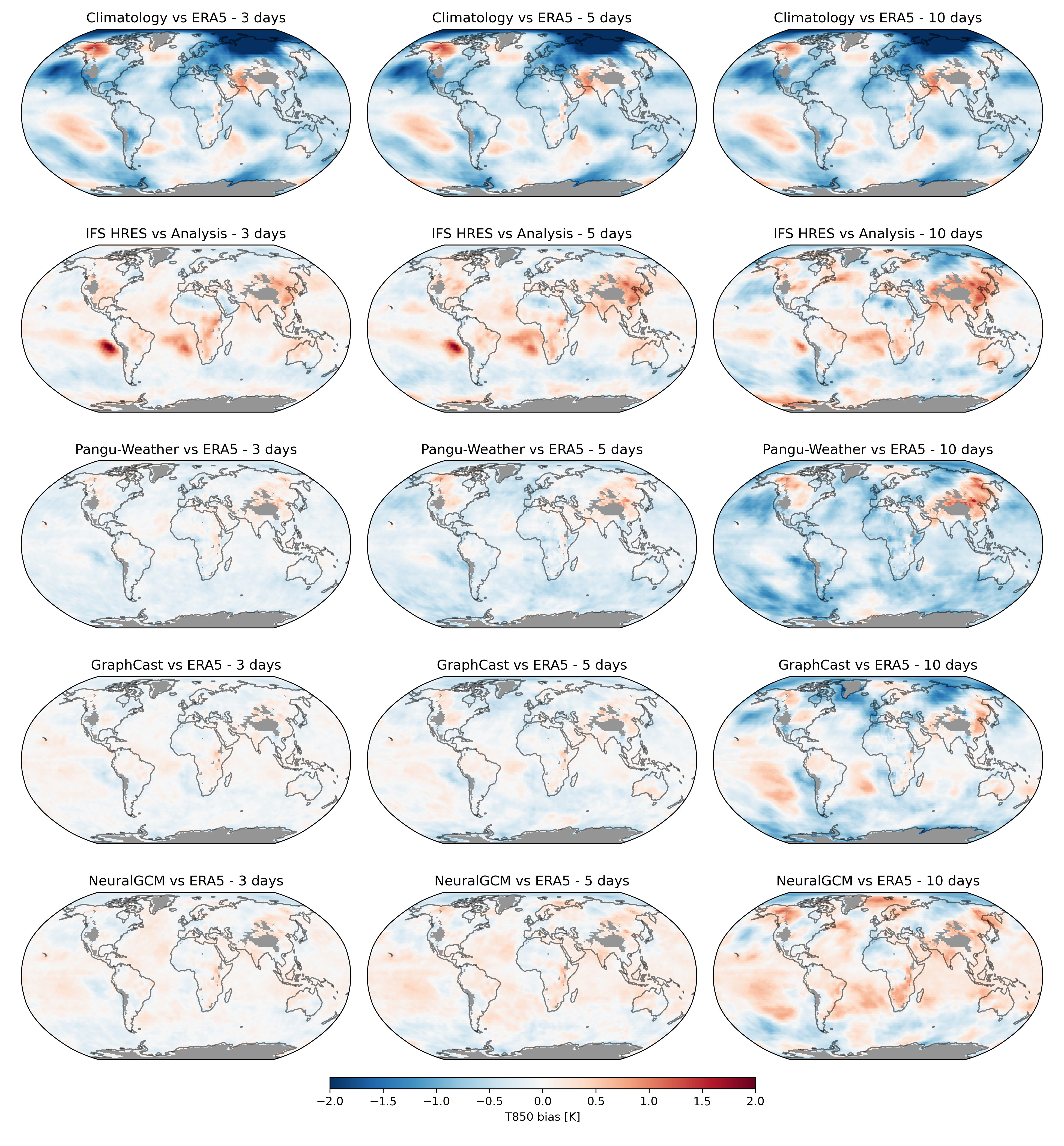}
\caption{Global mean 850hPa temperature bias for 3, 5 and 10 day lead times. Below ground areas are grayed out.}
\label{fig:bias_t850}
\end{figure}

\begin{figure}
\includegraphics[width=\textwidth]{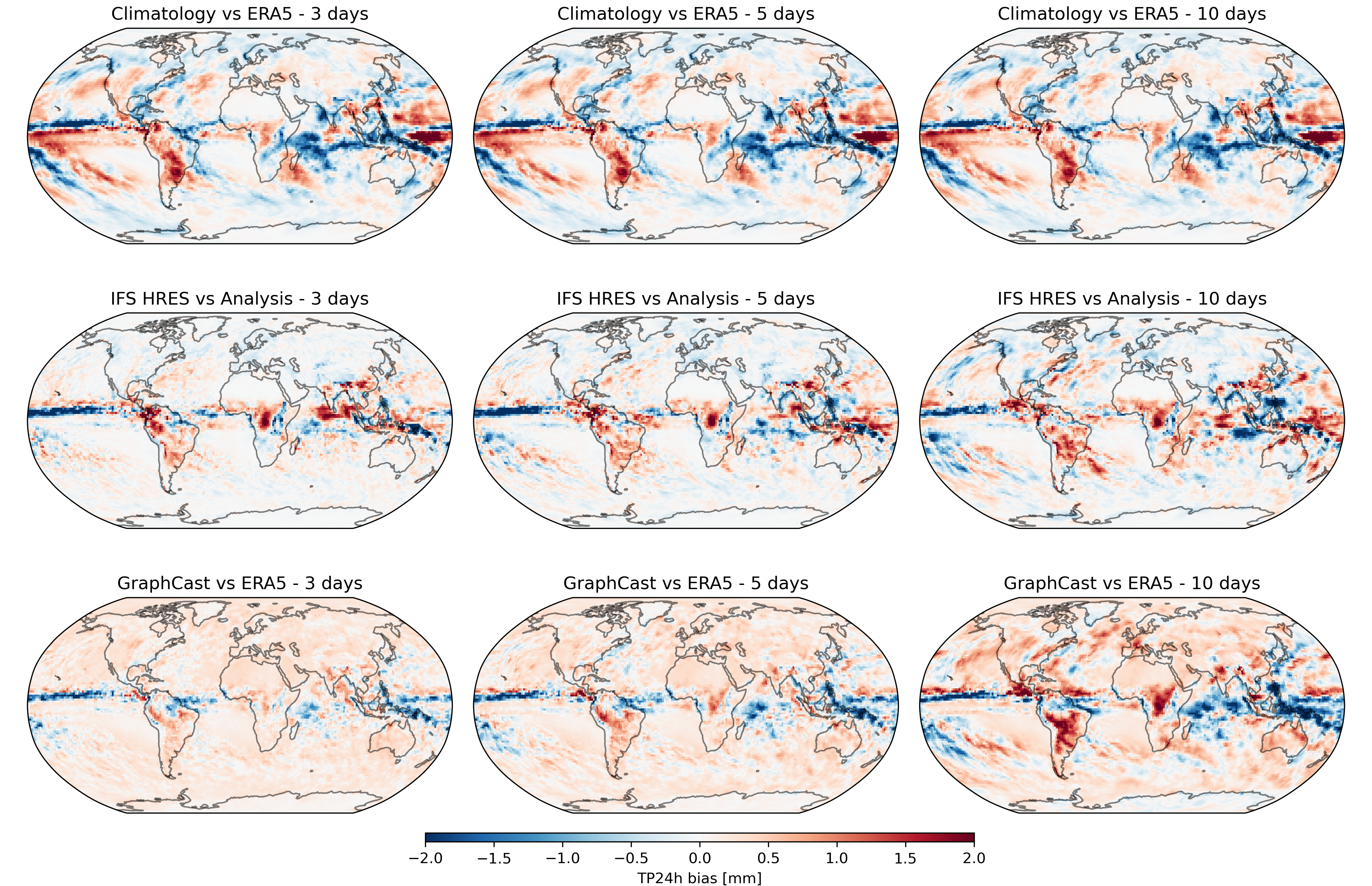}
\caption{Global mean 24h precipitation accumulation bias for 3, 5 and 10 day lead times.}
\label{fig:bias_tp24h}
\end{figure}

\begin{figure}
\includegraphics[width=\textwidth]{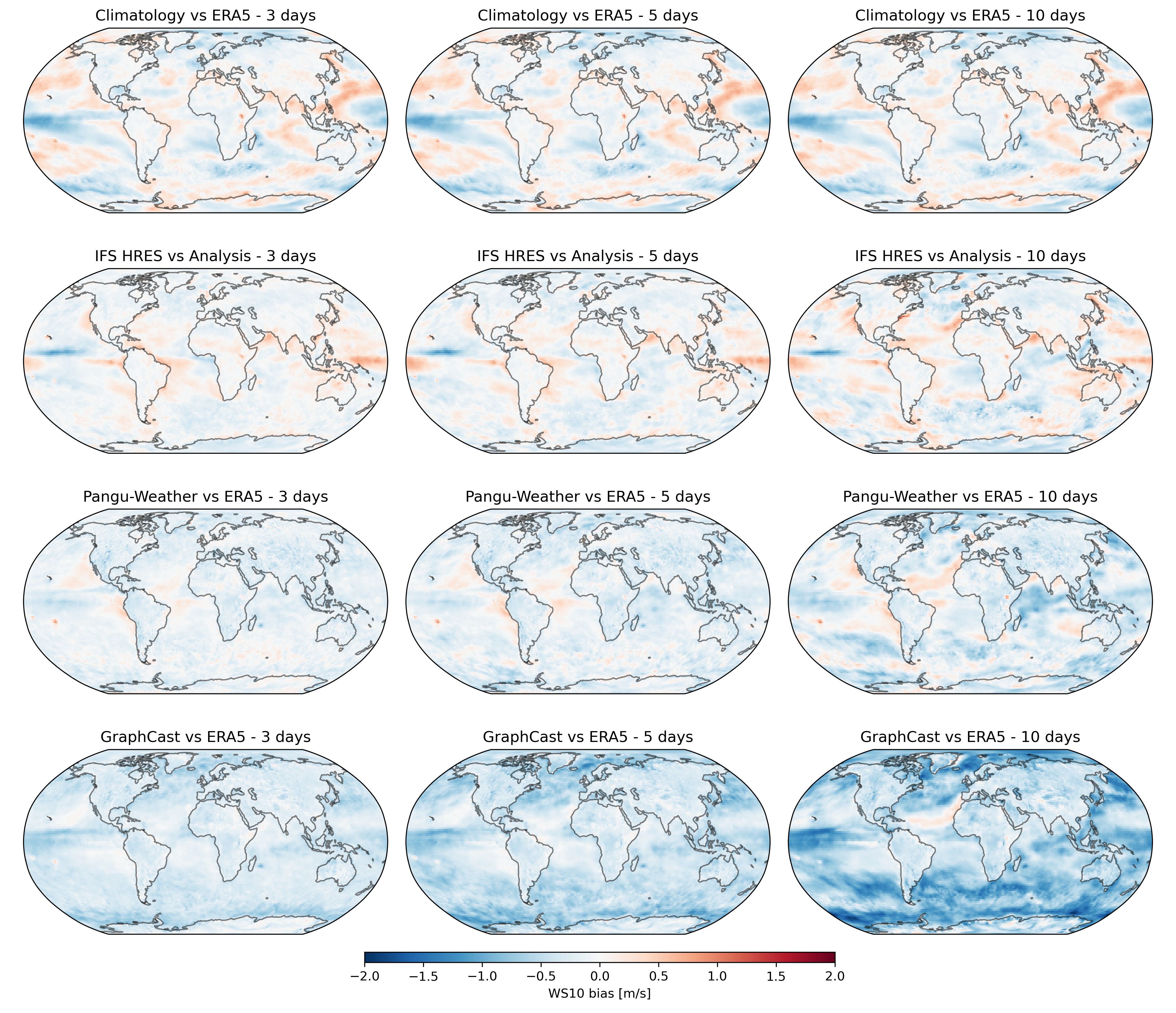}
\caption{Global mean 10m wind speed bias for 3, 5 and 10 day lead times.}
\label{fig:bias_ws10}
\end{figure}

\begin{figure}
\includegraphics[width=\textwidth]{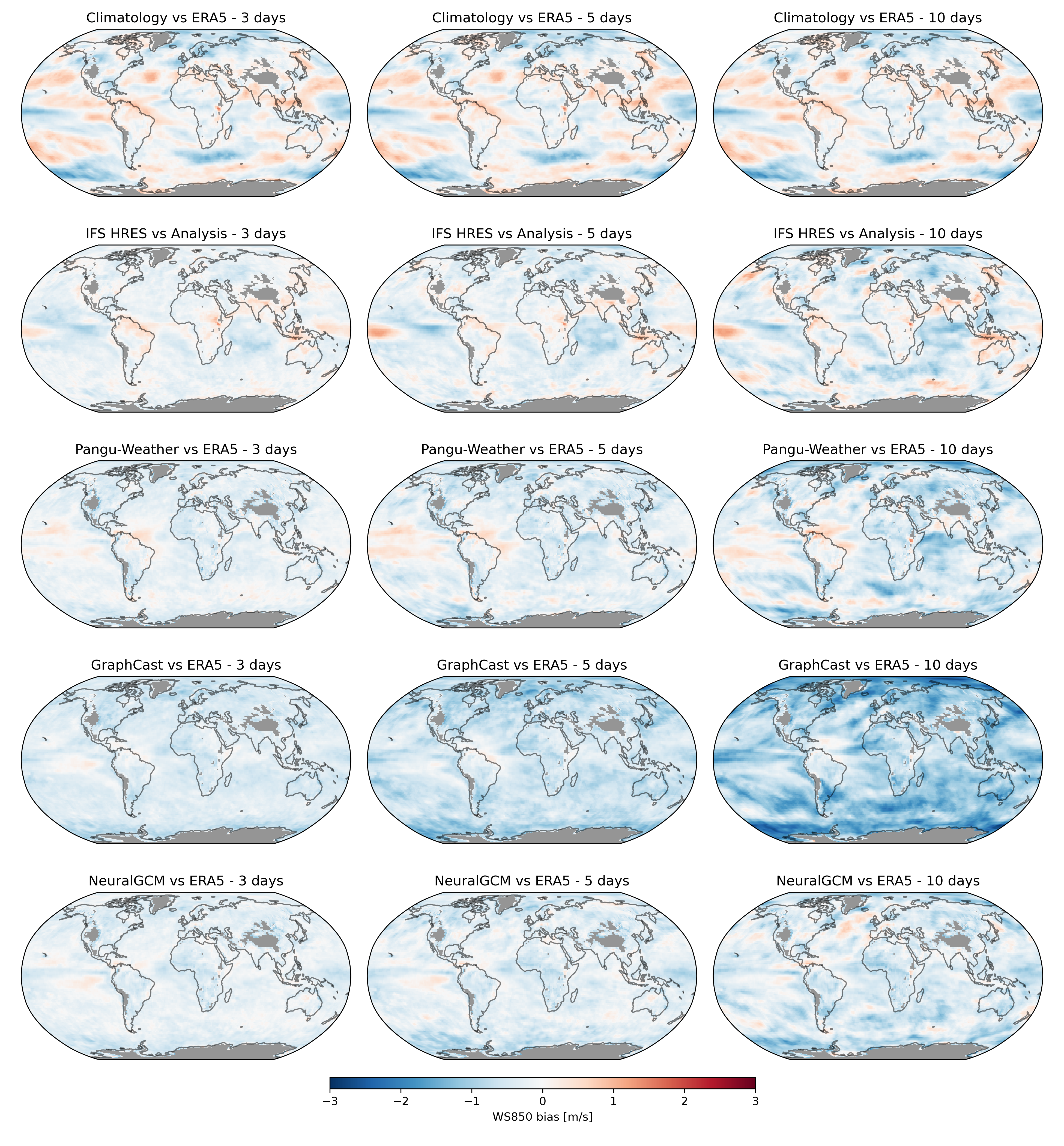}
\caption{Global mean 850hPa wind speed bias for 3, 5 and 10 day lead times. Below ground areas are grayed out.}
\label{fig:bias_ws850}
\end{figure}

\begin{figure}
\includegraphics[width=\textwidth]{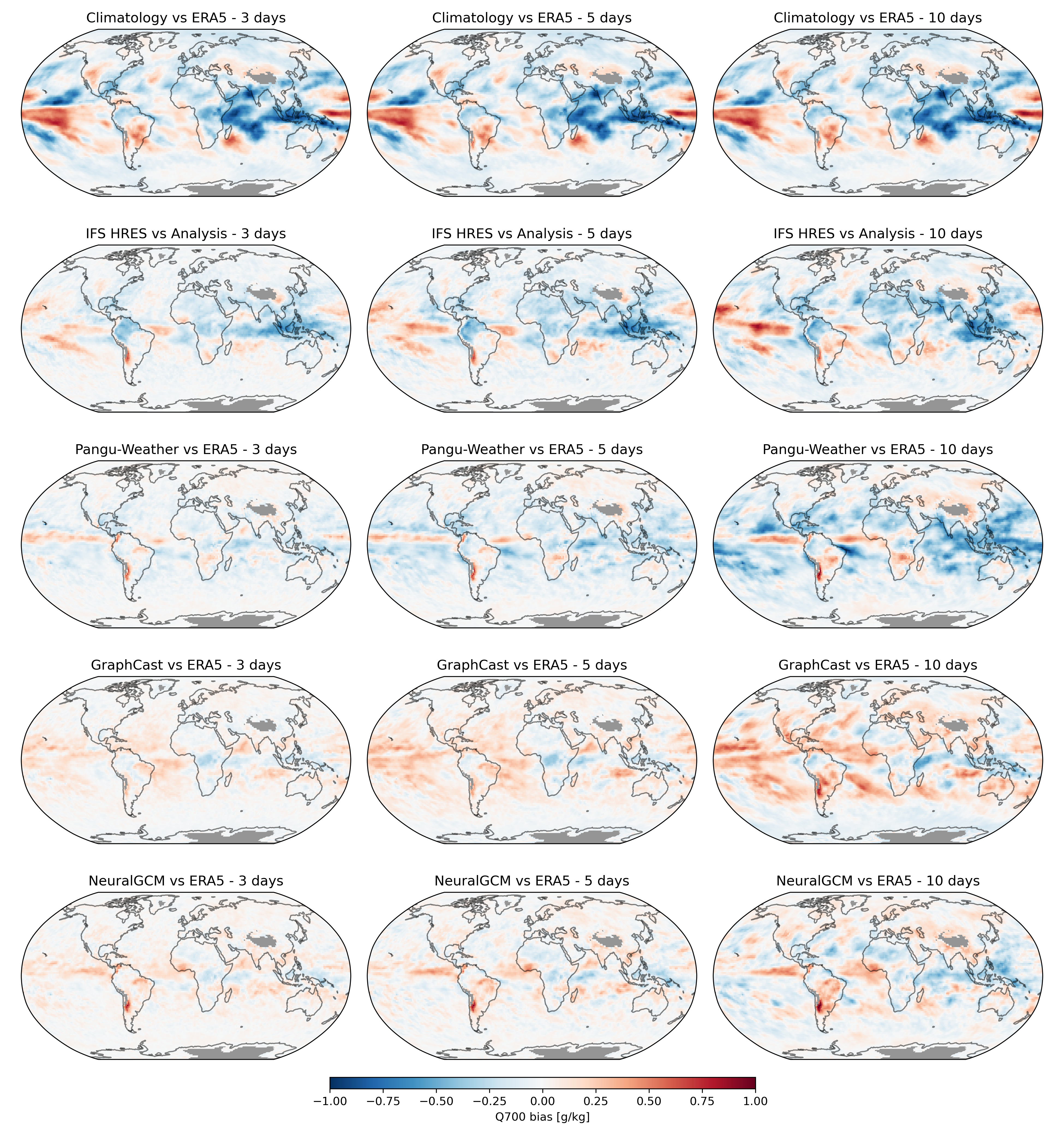}
\caption{Global mean 700hPa specific humidity bias for 3, 5 and 10 day lead times. Below ground areas are grayed out.}
\label{fig:bias_q700}
\end{figure}

\begin{figure}
\includegraphics[width=\textwidth]{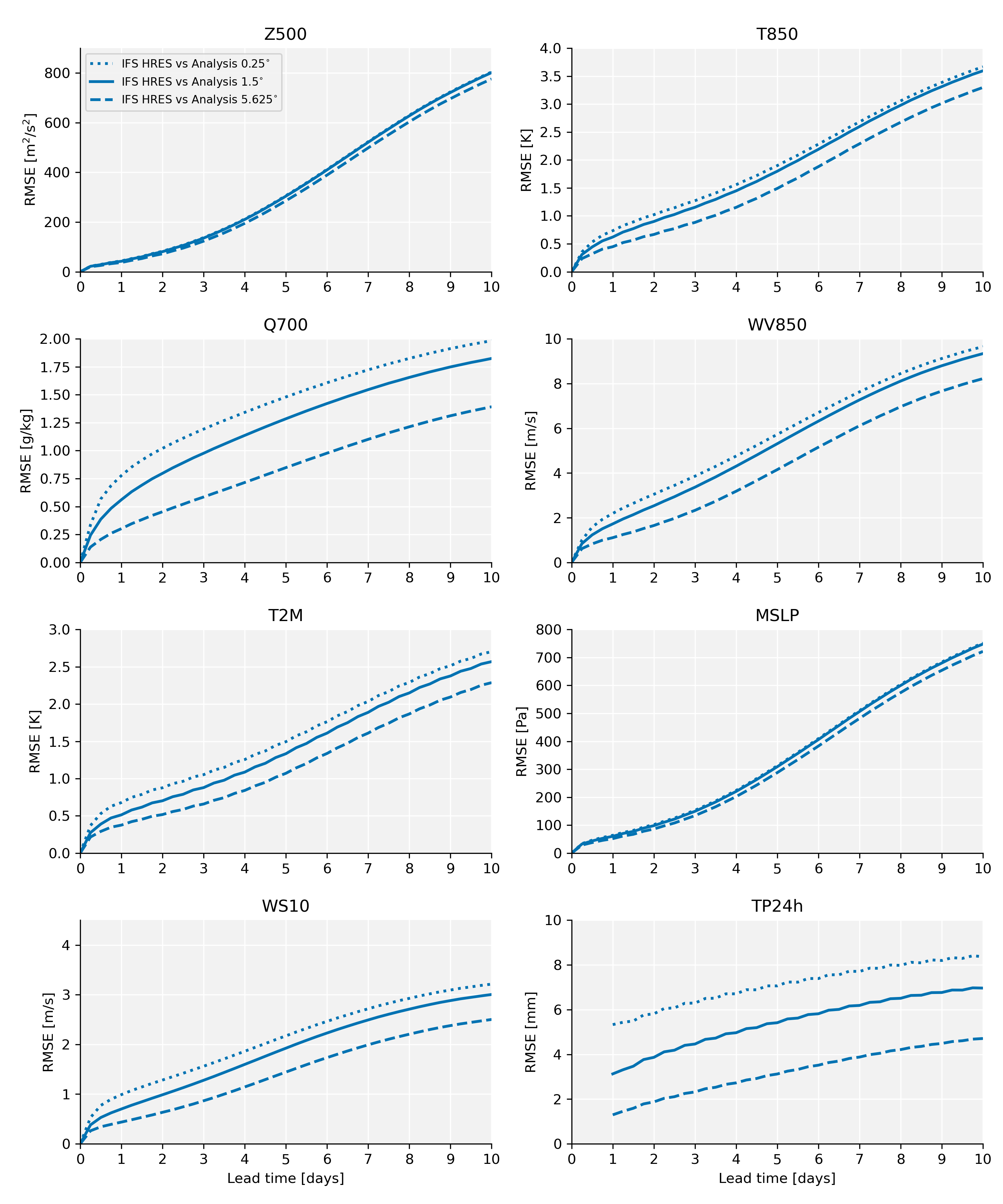}
\caption{Comparison of RMSE scores for IFS HRES evaluated at different resolutions}
\label{fig:resolution_IFS}
\end{figure}



\begin{figure}
\includegraphics[width=\textwidth]{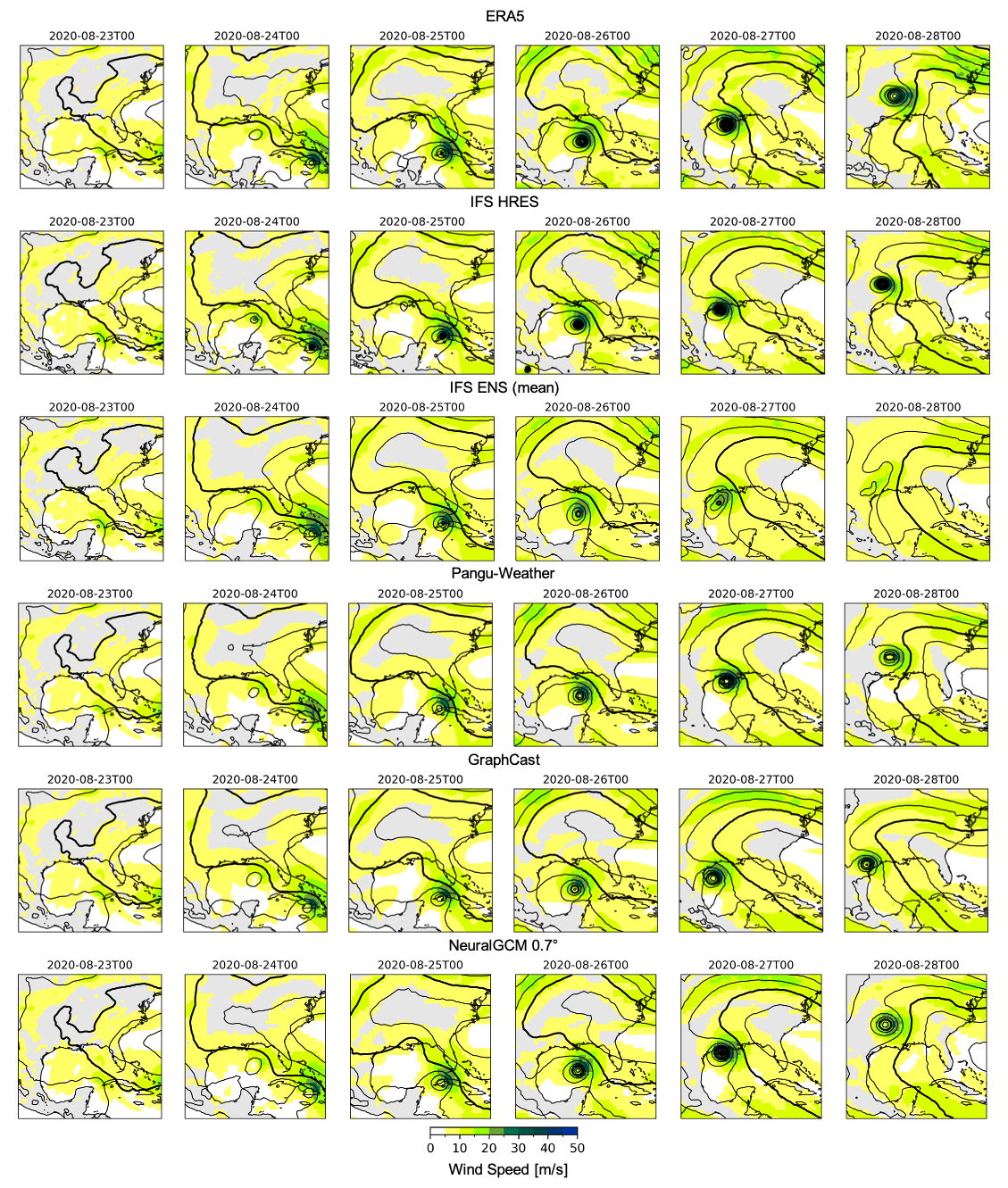}
\caption{Case study: Hurricane Laura. Top row shows ERA5 ground truth. Other rows show forecasts initialized at 2020-08-23 00UTC. Lines show 850hPa geopotential contours (not available in our Pangu-Weather dataset). Shading shows 850hPa wind speed.}
\label{fig:case_laura}
\end{figure}

\end{document}